\documentclass[manuscript=article]{achemso}
\SectionNumbersOn
\usepackage[version=3]{mhchem} 
\usepackage[utf8]{inputenc}
\usepackage[T1]{fontenc}
\usepackage{graphicx}
\usepackage{amsmath}
\usepackage{amssymb}
\usepackage{amsfonts}
\usepackage{bm}
\usepackage{dcolumn}
\usepackage{bm}
\usepackage{hyperref}
\usepackage[mathlines]{lineno}
\usepackage{xcolor}
\usepackage{titlecaps}
\usepackage{bbold}
\usepackage{color,soul}
\definecolor{sph}{rgb}{0.0588, 0.3216, 0.7294} 
\definecolor{ppk}{rgb}{1.0, 0.4549, 0.0902} 

\newcommand{\nc}{\newcommand}
\nc{\I}{$I$}
\nc{\II}{$II$}
\nc{\III}{$III$}
\nc{\nn}{\nonumber}
\def\e{\mathcal{E}}

\nc{\XYZ }{\bf}
\nc{\ABC }{\st}

\newcommand{\beq}{\begin{equation}}
\newcommand{\eeq}{\end{equation}}
\newcommand{\bessel}{\mathcal{J}}
\newcommand{\onlinecite}[1]{\citenum{#1}}

\newcommand{\beqar}{\begin{eqnarray}}
\newcommand{\eeqar}{\end{eqnarray}}

\def\beq{\begin{equation}}
\def\eeq{\end{equation}}
\def\beqa{\begin{eqnarray}}
\def\eeqa{\end{eqnarray}}
\def\e{\varepsilon}

\def\ve{\varepsilon}

\def\cH{{\mathcal H}}

\def\al{\alpha}

\def\si{\sigma}

\nc{\YD}[1]{{\color{red}{{#1}}}}
\nc{\AS}[1]{{\color{Magenta}{{#1}}}}
\nc{\OLA}[1]{{\color{Green}{{#1}}}}
\nc{\subhajit}[1]{{\color{sph}{{#1}}}}
\nc{\rs}[1]{{\color{black}{{#1}}}}
\author{Subhajit Sarkar}
\email{subhajit.sarkar@snu.edu.in}
\affiliation[SNU]{Department of Physics, School of Natural Sciences, Shiv Nadar Institution of Eminence (SNIOE), Deemed to be University, NH91, Tehsil Dadri, Gautam Buddha Nagar, Uttar Pradesh -- 201314, India}

\author{Amos Sharoni}
\email{amos.sharoni@biu.ac.il}
\affiliation[BGU]
{Department of Physics, Bar Ilan University, Ramat Gan, Israel}

\author{Oliver L. A. Monti}
\email{monti@arizona.edu}
\affiliation[UoA]
{Department of Chemistry and Biochemistry, University of Arizona, 1306 E. University Blvd., Tucson, Arizona 85721, United States \\Department of Physics, University of Arizona, 1118 E. Fourth Street, Tucson, Arizona 85721, United States}

\author{Yonatan Dubi}
\email{jdubi@bgu.ac.il}
\affiliation[BGU]
{Department of Chemistry, Ben Gurion University of the Negev, 1 Ben-Gurion Ave, Beer Sheva, 8410501, Israel}

\title{The spinterface mechanism for the chiral-induced spin selectivity effect: A Critical Perspective}

\abbreviations{CISS}

\begin{document}
\begin{abstract}
\begin{figure*}
\begin{center}
\includegraphics[width=7.6cm]{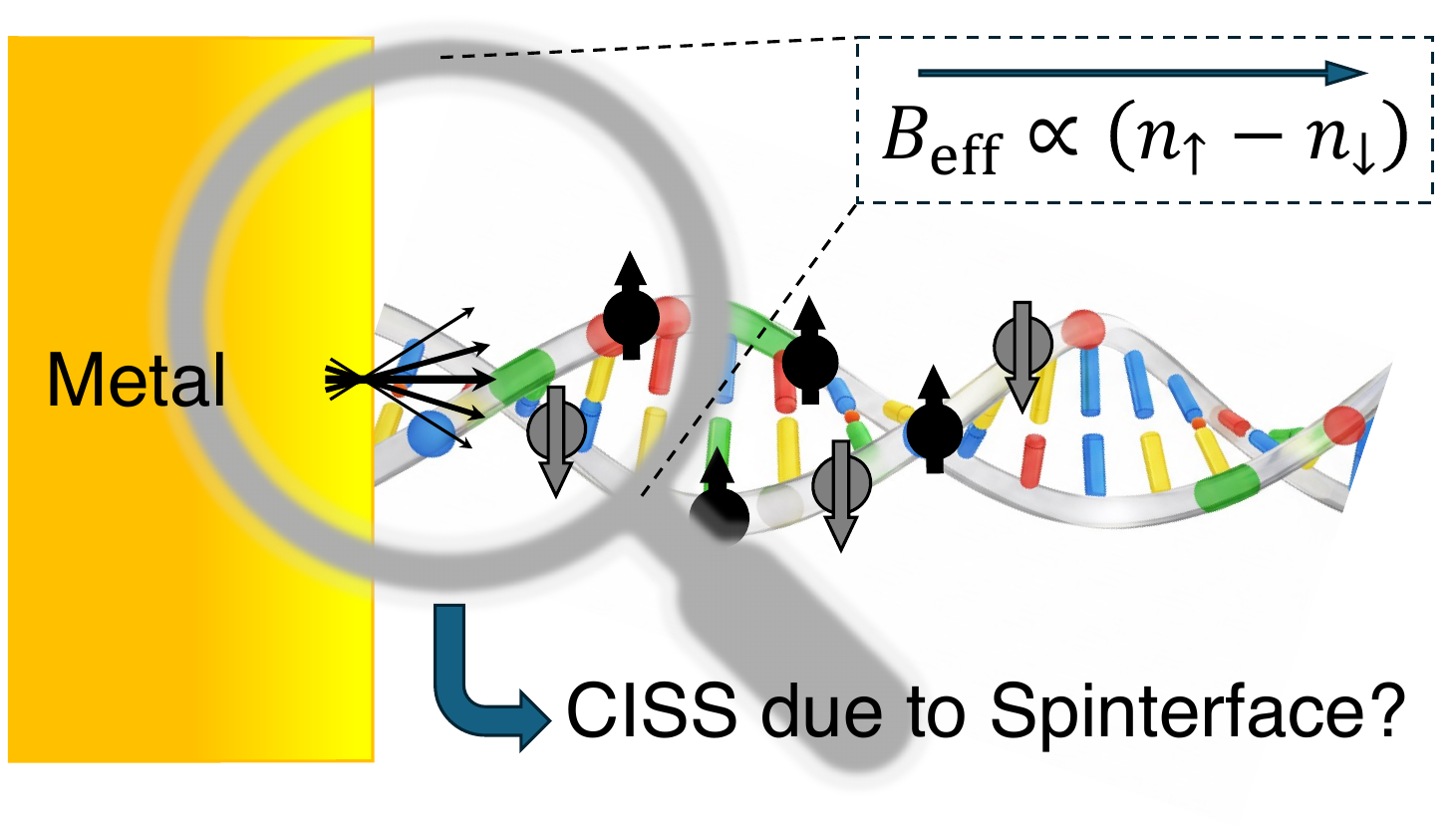}
\end{center}
\end{figure*}
The chiral-induced spin selectivity (CISS) effect, whereby chiral molecules preferentially transmit electrons of one spin orientation, remains one of the most intriguing and debated phenomena at the interface of spintronics, molecular electronics, and quantum materials. Despite extensive experimental observations across diverse platforms - including transport junctions, photoemission, and enantioselective chemistry - a comprehensive theoretical framework is still lacking. In this perspective, we critically examine the spinterface mechanism as a unifying explanation for the CISS effect. The spinterface model, which hypothesizes a feedback interaction between electron motion in chiral molecules and fluctuating surface magnetic moments, is shown to quantitatively reproduce experimental data across various systems and conditions. We contrast it with some existing theoretical models, highlighting key experimental features. Importantly, we also address open questions and criticisms of this model, including the nature of surface magnetism, the role of dissipation, and the applicability of the mechanism to non-helical or electrode-free systems. By offering falsifiable predictions and reconciling theory with experimental raw data, this work aims to sharpen the dialogue surrounding the microscopic origin of CISS and stimulate further experimental and theoretical progress.

\end{abstract}
\section{Introduction}
The chirality-induced spin selectivity (CISS) effect, discovered more than two decades ago \cite{Ray99}, is the general name of a set of experimental findings that reveal that chiral molecules can preferentially transmit electrons with a specific spin orientation. This selectivity is not due to an external magnetic field but rather to intrinsic properties of the system, most notably the chirality of the molecules, indicated by the fact that the preferred electron spin orientation is reversed when the molecular chirality is reversed. 

Interest in the CISS effect has boomed in recent years for two central reasons. First, it is a potentially useful effect that can be harnessed for a variety of technologically relevant applications. Some examples include spintronic devices (chiral-molecule-based spin-valves and spin-filters, chiral-based magnetic memory devices, and sources of spin-polarized electrons) \cite{Bloom2024chiral,Naaman12, kim2021chiral}, chemistry and catalysis (enantiomer separation, spin-dependent and enantio-selective chemistry)~\cite{bloom2024chemical}, material science (chiral hybrid organic-inorganic materials for spin filtering, chiral semiconductors, and semimetals for spintronics) \cite{waldeck2021spinmaterials}, energy conversion (spin-selective solar cells, chiral-based thermoelectric devices, and spin-dependent electrodes for batteries and fuel cells) \cite{bloom2024chemical,Bloom2024chiral}, and even quantum technologies (quantum spin-state preparation and manipulation and chiral-based qubits and quantum gates). \cite{aiello2022chirality,chiesa2021assessing,Bloom2024chiral}. 

The second reason is that, although the CISS effect has now been known for more than a quarter of a century, there is still no agreement on the origin of this effect. Since its discovery, many theoretical postulates have been put forward to explain the CISS effect \cite{evers2022theory}, yet despite this theoretical effort, no consensus has emerged. This is further complicated by the fact that many of the experimental features also remain unclear. This has been recognized by the CISS community. A recent example is the review by Evers et al. \cite{Evers_RevModPhys}, where the CISS effect takes up a considerable portion of the ``Open Questions” section. The authors write: ``At present, it appears that a large gap remains between the experimental observations and the quantitative estimates from theory." The knowledge gap seems to be very basic - at its heart, the origin of the CISS effect is largely unknown, and there is an ongoing debate on what even constitutes the CISS effect and its signatures.

The goal of this perspective is to summarize the approach known as the "spinterface model", put forward in a series of papers \cite{alwan2021spinterface, dubi2022spinterface, yang2023, alwan2023temperature, alwan2024role,monti2024surface}, as a plausible explanation for the CISS effect. However, we wish to go beyond a standard review and critically discuss the spinterface model (and other theories), by confronting it with existing criticism, and by providing a series of old and new predictions which arise from the spinterface model, which we invite the experimental CISS community to test and the theory community to challenge. 

We organize this perspective as follows. We start with a brief and necessarily somewhat incomplete review of the experimental side of the CISS effect (Sec.~II). We then briefly review some of the main theories for the CISS effect (Sec.~III), and outline various difficulties that arise with each of them. Next, we introduce the spinterface model and its remarkable success in explaining experiments and reproducing experimental {\it raw} data (Sec.~IV). We then outline the main criticism of the spinterface model and discuss our current thoughts on a response to these claims (Sec.~V). In Section~VI, we provide a list of old and new predictions of the spinterface model. And finally, we summarize and conclude.


\section{A brief review of the main experimental facts}\label{brief review}

Since the literature on the CISS effect is overwhelmed by a surprisingly large number of review articles, we take the liberty here of giving only a brief overview, highlighting what we believe are the essential experimental ingredients of the CISS effect. We differentiate between different experimental manifestations of the CISS effect, and at the end of the section, enumerate what are the common features.  

The first demonstrated manifestation of the CISS effect is in photoemission  \cite{mollers2022chirality, kettner2018chirality, mollers2022spin, Gohler11, abendroth2019spin, badala2022vectorial, nino2014enantiospecific}. In photoemission CISS experiments, electrons are photoexcited out of a non-magnetic metal surface covered with chiral molecules. The photoemitted electrons emerging from the surface exhibit some degree of spin polarization, which depends on the handedness of the chiral molecules. 

The by far more experimentally common manifestation of the CISS effect is the transport CISS effect (T-CISS) \cite{Naaman12, Naaman15, naaman2019chiral, naaman2020chiral}, where electric current flows through a junction composed of a chiral molecule sandwiched between a metallic and a ferromagnetic electrode. The current-voltage characteristics of this chiral-molecular junction vary depending on whether the ferromagnet is magnetized parallel or anti-parallel to the molecular chiral axis, which is also nearly the axis of current flow. \rs{While many experiments use a collinear geometry, transverse CISS—where magnetization and/or current are perpendicular to the chiral axis—has also been reported, and such scenarios are indeed regarded as part of the broader phenomenology \cite{wang2024transverse}.} Importantly, as Fransson stressed \cite{fransson2023chiral}, what is measured in the T-CISS experiment is not the ``spin current,'' but rather the {\sl total} current, a distinction that is often overlooked. Equally important, what is observed as a signature of CISS is not ``spin polarization'' but rather magnetoresistance \cite{liu2023spin}.

While CISS is assumed to be the same effect for both transport and photoemission studies, significant distinctions exist between these phenomena, which may necessitate separate explanatory frameworks. For instance, transport studies examine non-equilibrium conditions and involve ferromagnetic electrodes, inherently breaking time-reversal symmetry in the system. In this context, understanding CISS requires investigating how chiral molecules affect spin-dependent transmission in nanoscale devices. In contrast, there is no apparent external agent to break time-reversal symmetry in photoemission experiments. \rs{Furthermore, transport experiments implicitly define a preferred direction; the transport axis is the one that often aligns with the chiral axis of the molecule that is almost always helical, while transverse arrangements can also yield robust signals \cite{wang2024transverse}}. In photoemission experiments, no such preferred direction is defined, at least if one were to measure photoelectrons emerging at all possible angles with respect to the surface normal. Such subtleties are essential when constructing theoretical models for different CISS manifestations. 

While these are the two central manifestations of the CISS effect, they are by no means the only ones. For instance, magnetic measurements of chiral-molecule-decorated metal surfaces have been reported to show permanent magnetization despite the absence of an external field \cite{ben2017magnetization,koplovitz2019single,metzger2020electron,tassinari2018chirality,abendroth2019spin, abendroth2017analyzing,Bloom2024chiral,mishra2024inducing,theiler2023detection,nguyen2024mechanism}. Other examples - many listed and described in detail in the many other reviews - include enantio-chemistry, enantio-electrochemistry, CISS in spintronic devices, etc. \cite{Bloom2024chiral}. 

What then are the most essential features of the CISS effect, which are common to all CISS manifestations? The first and foremost is the dependence on molecular chirality. In all experiments, the CISS signal, be it current, photoemission intensity, reaction rate, or any other measurable quantity, is different for the two chiral enantiomers. This measured quantity can be used to define a majority and minority quantity, often associated with majority and minority spins. For instance, in T-CISS, the direction of the ferromagnetic (FM) electrode along which the current is higher is defined as the majority spin direction. Note that, clearly, both spins participate in transport. Most importantly, in all instances of the CISS effect, when the chirality of the molecule is flipped, the direction of the majority quantity is also flipped. This is a crucial point, because it points to some fundamental symmetry of the effect, as recently discussed by Rikken and Avarvari \cite{rikken2023comparing}: In the T-CISS effect, the majority species depends on chirality but not on the direction of current flow, a distinct symmetry which is different from other chirality-related effects such as magneto-chiral anisotropy.  Thus, flipping the molecular chirality is a crucial control for any experiment that claims to show a manifestation of the CISS effect. 

Another crucial feature of the CISS effect is its generality, more specifically, its appearance on a very wide scale of systems. The T-CISS, for instance, has been measured in small single molecule junctions \cite{aragones2022magnetoresistive,yang2023}, intermediate-size molecules such as helicene, large bio-molecule such as polypeptides and oligonucleotides (i.e. short, single- or double-stranded DNA or RNA)  \cite{bloom2024chemical}, large chiral supra-molecular structures \cite{garcia2021importance,mondal2021spin} and giant Polyaniline (PANI) structures \cite{jia2020efficient,Bloom2024chiral}, as well as in layers of chiral solid materials\cite{bloom2024chemical,Bloom2024chiral}. This is a strong requirement if CISS is a fundamental effect. 

A universal feature of almost all reported CISS observations thus far is the use of a metal electrode. With a few exceptions (which will be discussed in Sec.\ref{critical}), all current CISS experiments involve some sort of metallic or sometimes semi-metallic electrode. In T-CISS, the metal electrode is part of the junction, while in photoemission CISS experiments and magnetization measurements it is the substrate on which chiral molecules are placed. The latter is also the case for CISS enantio-chemistry and enantio-electrochemistry. To date, it is unclear whether metallic substrates are required for the CISS effect to manifest or maybe simply necessary for it to be observed.

One of the most frequently used measures for the strength of the CISS effect in transport measurements is the so-called CISS polarization, $P_{CISS}$, defined as the ratio of the difference between the currents measured for the two FM electrode magnetization directions and the total current (i.e., their sum), $P_{CISS}=\frac{J_+ - J_-}{J_+ + J_-}$, where $J_\pm$ refers to the current when the FM electrode is polarized parallel/antiparallel to the direction of current flow. We note that a better phrase for this measured parameter would be CISS {\sl magnetoresistance}, yet we use them here interchangeably because the former is often used in the literature. \rs{Importantly, $P_{CISS}$ should not be confused with `standard' magnetoresistance, which follows from the common definition in spintronics, and in the CISS context is evaluated as $MR_{CISS}=\Delta MR= \frac{J_+ - J_-}{J_0}$ and is not bound from above.}

An important experimental observation from T-CISS experiments is that the CISS polarization can be extremely high, reaching $>90\%$. The high value is in itself already a surprisingly impressive result. It is even more surprising that this is the case in experiments where the FM electrode has finite spin polarization, such as is, e.g., the case for Ni (maximal spin polarization of $\sim 30\%$), used in the vast majority of T-CISS experiments. From the perspective of a Julli\`ere model of magnetoresistance and if the CISS effect is a result of spin-filtering by the molecule alone, the CISS magnetoresistance ratio cannot exceed that of the Ni spin polarization. \rs{When interactions are negligible or weak, the description, in terms of the Julli\`ere model, is valid and the CISS polarization is bounded by the Ni spin polarization \cite{alwan2024role}.}
This provides a crucial hint on how the nature of the CISS effect differs fundamentally from more standard magnetotransport effects. 

In contrast to these universal aspects, there are a variety of experimental features that seem not to be so universal. Two examples are the dependence of CISS on the length of the chiral molecules and on temperature. While early measurements seemed to show that the CISS polarization increases with molecular length, various later measurements showed no distinguishable length dependence (see, e.g., \cite{Xie11} and discussion in \cite{dubi2022spinterface}). Similarly, different experiments show that the CISS effect either increases \cite{das2022temperature},  decreases \cite{alwan2023temperature}, or remains unchanged \cite{mollers2024probing} with increasing temperature.

The take-home message from this brief description of CISS experiments (a far more detailed review of experimental techniques is provided in Ref.~\onlinecite{Bloom2024chiral} as well as in numerous other reviews) is as follows:  when an experimental paper on the CISS effect is read, one should be careful to distinguish between universal and experiment-specific CISS features. Specifically, one should be careful in making sweeping generalizations of CISS features based on specific experiments. As seen in the case of temperature- and length-dependence, there are various, and sometimes conflicting,  examples for many features of the CISS effect. 

The point of distinguishing between universal and system-specific CISS features is, in fact, critical when we come to assess the success of different theoretical descriptions of the effect. After all, the role of theory in this field is at least in part to explain experimental findings, and one critical way of doing so is by comparing results of models with results from experiments (e.g. \cite{frigg2006models,frigg2020modelling,frigg2022models}). Both theorists and experimentalists should thus be very aware whether an observed feature is universal and hence should be there for every specific system considered under the model, or whether it is system-specific, and hence the model should allow for different outcomes for this feature. Put simply, a convincing theory for the CISS effect should always show the universal features and should be flexible enough to accommodate different manifestations of system-specific features, since only then can its validity be properly evaluated. 

\section{A brief review of CISS theory}\label{review_theory}
While not as extensively reviewed as the experimental aspects of the CISS effect, different theories for the CISS effect were also recently reviewed \cite{evers2022theory,Bloom2024chiral}. We thus, again, keep this part of the perspective short and discuss only those select theoretical approaches which we believe are relevant to the discussion. Readers are encouraged to read the more detailed reviews of the theory for a broader appreciation of the work in the field.

\subsection{What is the theoretical difficulty?}\label{What is the theoretical difficulty?}
A spin-dependent transport effect, which is inherently related to the molecular chirality, can be described with a very simple single-level transport toy model (Breit-Wigner model). As is common in quantum transport, consider the molecular junction as consisting of a single level, representing the molecular frontier orbital, coupled to metallic (left) and FM (right) electrodes. Then, under voltage bias a current flows. If the molecule is helical, then electrons traveling through the molecule can be viewed as traveling through a tiny solenoid, and from the Faraday-Lentz law, the current will generate a magnetic field. This field interacts with the electron spins, generating an effective current-induced Zeeman splitting of the molecular orbital, which leads to different I-V curves for two spins. 

Let us put this toy model into mathematical form. We use the standard Landauer formula in the wide-band approximation \cite{Book:Cuevas_Scheer10,Di_ventra08} to describe transport. Assume that the spin-direction $\sigma$ can be parallel or anti-parallel to the current flow, and that $s=\pm1$ represent the magnetization of the FM electrode parallel or anti-parallel to the current axis (see discussion in \cite{fransson2023chiral}). Then, the metal electrode is characterized by a constant level broadening parameter $\Gamma$, and the FM electrode has a level broadening $\Gamma_{\si s}$ which depends on both $\sigma$ and $s$. For the sake of simplicity we set it to $\Gamma_{\si s}=\Gamma$ if $\si=s$ and $\Gamma_{\si s}=0.5 \Gamma $ if $\si\neq s$, representing an FM spin polarization of $\sim30\%$ \cite{alwan2024role}. 

The total current through the device, for different FM directions $s$, is thus given by 

\beq
J_s=\frac{e}{h} \sum_\si \int dE \frac{\Gamma \Gamma_{\si s}}{\Gamma \Gamma_{\si s}+(E-\varepsilon_\si)^2}\left(f_L(E-V/2)-f_R(E+V/2) \right)~~, \label{eq:toymodel}\eeq
where $f_{L,R}$ are the Fermi functions of the metal and FM electrodes, $V$ is the voltage, and $\varepsilon_\si$ is the spin-dependent orbital energy level. The solenoid field shifts the energy level through a Zeeman term of the form 
\beq
\e_{\si}=\e_0+\si \al J_s~~,
\label{eq:e_shift}\eeq
where $\e_0$ is the bare energy level in the absence of current, $J_s$ is the current in the presence of an FM electrode that is magnetized in the direction $s$, and $\al$ is the coefficient that converts the current to a shift in orbital energy. The sign of $\al$ determines the chirality (handedness) of the molecule, since it is determined by the handedness of the solenoid. Since $J_s$ appears in the integrand itself, Eqs.~(\ref{eq:toymodel}-\ref{eq:e_shift}) are to be evaluated self-consistently. 

\begin{figure}
\begin{center}
\includegraphics[width=12.6cm]{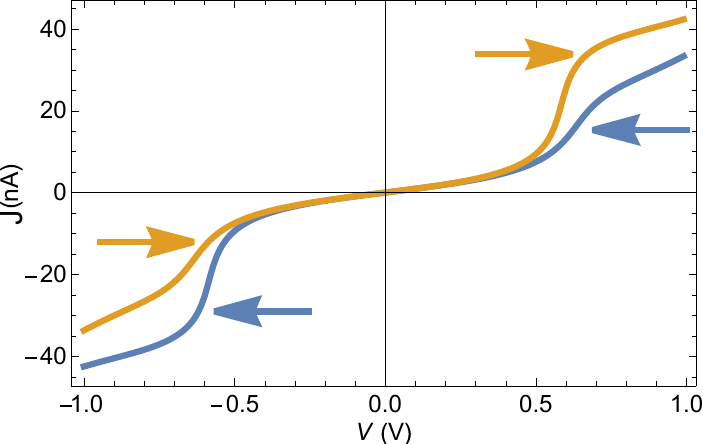}
\caption{ I-V curves -- current $J$ in nA as a function of voltage $V$ -- for the two FM magnetization directions for the self-consistent toy model, Eqs.~(\ref{eq:toymodel}-\ref{eq:e_shift}). Electrons move from left to right.}
\label{fig:IV-toymodel}
\end{center}
\end{figure}

In Fig.~\ref{fig:IV-toymodel} we plot the I-V curves for the two magnetizations pointing parallel and anti-parallel to the direction of current flow (other numerical parameters are $\Gamma=5$meV, $\e_0=-0.4$ eV, temperature $T=300$ K and $\al=5\times 10^{-3}$eV). The results do not depend qualitatively on the choice of these parameters. Rewardingly, this simple toy model leads indeed to different I-V curves for different FM magnetization direction. 

However, there are two critical flaws of this model, which cannot be amended easily by simply introducing increased sophistication to the transport description. First, notice that the {\sl symmetry} of the I-V curves is wrong, because what is the majority direction for positive voltages (orange line, showing higher current) turns to be the {\sl minority} direction at negative voltages. This symmetry agrees with the so-called magneto-chiral anisotropy (MChA) effect \cite{atzori2021magneto}, but not with the CISS effect \cite{rikken2023comparing}. This is a crucial observation, because the symmetry is a universal feature of different magnetoelectric effects: While the simple explanation presented above can be relevant to observations of the MChA effect in molecular junctions, it does not seem to point to the CISS effect. Simply put, the CISS effect requires a symmetric dependence of the spin-dependent energetics on voltage, i.e. a version of Eq.~(\ref{eq:e_shift}) of the form $\e_{\si}=\e_0+\si \Lambda(V)$, with $\Lambda(V)=\Lambda_0+\Lambda_1 V^2$, and without a linear term.  

Yet another major difficulty with this simple model is that the magnitude of $\al$ is too large. One can easily evaluate $\al$ by noting that the solenoid field is $B_{sol}=\mu_0 n_t J$, where $\mu_0$ is the vacuum permeability ($4 \pi \times 10^{-7}$ T m/A), $n_t$ is the number of turns per unit length (in DNA, for example, $\sim 0.3$ (nm)$^{-1}$), and $J$, the current, and this should be multiplied by $0.5 g \mu_B$ for an electron, where $g$ is the electron g-factor and $\mu_B\simeq 5.788 \times 10^{-5}$ eV T$^{-1}$. The resulting solenoid field is about 7 orders of magnitude smaller than the value of $\al$ needed to detect a difference in the I-V curves for the two magnetization directions.  

This orders-of-magnitude discrepancy was detected early on (see, e.g. \cite{DiVentra11}). In an attempt to overcome it, a quantum version of the solenoid field effect was employed, namely the addition of spin-orbit interactions in the molecule (see, e.g., Eq.~3 in \cite{Ghazaryan20}). However, this cannot solve the problem because  spin-orbit coupling (SOC) in mostly carbonaceous molecules is too small, and a renormalization of $\sim 6$ orders of magnitude is still required in order to reproduce the experimental CISS polarization \cite{Ghazaryan20}. Enhancement of the SOC due to orbital curvature effects in the molecule (see, e.g., \cite{geyer2019chirality}) may increase it, but remain still far from presenting a possible solution to the order of magnitude puzzle. \cite{evers2022theory} We insist that any reasonable theoretical description of the CISS effect must be able to capture the magnitude of the effect with realistic physical parameters. We consider this an essential element of any theory of CISS.

An additional basic difficulty of CISS theories is the issue of time-reversal symmetry breaking. The appearance of spin polarization in photoemission experiments and magnetoresistance in transport suggests fundamentally the breaking of time reversal symmetry \cite{monti2024surface,evers2022theory,mena2024minimal}. Yet, at least on the face of it, transport and photoemission experiments appear time-reversal symmetric, and thus spin polarization and magnetoresistance should be prohibited, in contradiction with all reports of CISS. This difficult state of affairs is also true for most theoretical treatments that preserve time reversal symmetry. Any theory of CISS should address the mechanism of time reversal symmetry breaking.

\subsection{Examples of different theoretical approaches}

Before we continue to the spinterface model, we wish to give here two examples of existing theories for the CISS effect. The goal is to highlight the differences between the different theoretical approaches and to articulate some of the critique of these models. We limit the discussion here to only two examples, and refer the interested reader to the recent review of CISS theory by Evers and co-authors \cite{evers2022theory}.  

Guo and Sun proposed a model \cite{Guo12a}, extended further later \cite{Guo12b,Guo14a,pan2016spin}, based on a tight-binding description of transport. It takes into account (i) the double-stranded (or multiple pathway) nature of the molecule, (ii) (discretized) spin-orbit interactions along the molecule, (iii) dephasing processes through Buttiker probes \cite{buttiker1985small,buttiker1986role}, with an understanding that a combination of all three is required to generate spin polarization in a two-terminal device \cite{evers2022theory}. Within this model, the authors were able to demonstrate substantial spin polarization, defined as the ratio \rs{$P_s=\frac{G_{\uparrow}(V)-G_{\downarrow}(V)}{G_{\uparrow}(V)+G_{\downarrow}(V)}$}, where $G_{\si}(V)$ is the voltage-dependent differential conductance in spin channel $\sigma$, in various systems such as DNA, Helicenes, and more. 

One immediate critique regarding this model and a number like it is that they are not evaluating the quantity measured in experiments. As described earlier \cite{fransson2023chiral,alwan2024role}, the experimental CISS polarization (or CISS magnetoresistance) is defined by measuring the {\sl total} current $J_m$ {\it for both spins} where the FM magnetization (denoted by the subscript $m$) is parallel ($m=+$ or $m=\uparrow$, the latter symbol is used frequently but is rather misleading) or antiparallel ($m=-$ or $m=\downarrow$). The CISS polarization is then defined as $P_{CISS}=\frac{J_+-J_-}{J_++J_-}$ (often referred to as the magnetoresistance ratio). The theory by Guo and Sun, on the other hand, is focused on calculating the differential conductances $G_{\si}(V)$ of two spin-channels and the polarization which ensues. As pointed out and demonstrated by Fransson \cite{fransson2023chiral}, the connection between these two quantities is nontrivial, and the CISS polarization cannot be simply deduced from the spin polarization. In fact, $P_{CISS}$  can have any value, even to vanish or be 100\%, even if $P_s$ is finite. In addition, these studies all considered zero-temperature systems, although the CISS effect has been measured across a broad range of temperatures, including at room temperature. 

A more serious criticism of this model is that it describes non-interacting electrons. In fact, the model by Guo and Sun is one example of a large family of models \cite{evers2022theory} which consider SOC in the molecule as the origin of the CISS effect, and approach the theoretical problem by describing the molecular system using a non-interacting Hamiltonian which includes SOC in various forms. However, we have recently shown \cite{alwan2024role} that a non-interacting approach cannot describe the CISS effect. The reason is that if the system is non-interacting, i.e. the Hamiltonian can be cast as a single-electron theory, then {\sl by definition} the CISS polarization is confined from above by the polarization of the ferromagnetic electrode, $P_{FM}$, as expected from the Julli\`ere formula.

To understand this general rule, analytically and numerically demonstrated in [\cite{alwan2024role}], consider the case that the molecule is a {\sl perfect} spin-filter. If the FM electrode is aligned such that its polarization is parallel to the direction of the spins allowed by the chiral molecule, it will have some characteristic I-V response. Now, if the FM magnetization is reversed, the same spins are allowed to pass the molecule (which did not change), but are now the minority spins of the FM. In the FM, the ratio between currents of minority and majority species is given by the FM polarization, $\lambda=\frac{J_{min}}{J_{maj}}=\frac{1-P_{FM}}{1+P_{FM}}$. This means that when the FM magnetization is reversed, there will still be a current $J_{min}=\lambda J_{maj}$, which will lead to CISS polarization equal to $P_{FM}$. If the molecule is not a perfect spin-filter, the resulting CISS polarization will necessarily be lower, and thus this argument provides an upper limit for the CISS polarization for non-interacting models. However, in numerous transport experiments \cite{alwan2024role} a CISS polarization much larger than that of the FM electrode is measured. Again, typically a Ni electrode is used which implies $P_{FM}\leq 35\%$. This means that all theories which consider transport through a chiral molecule using non-interacting electrodes are in qualitative disagreement with many experiments.

A different theoretical approach was suggested by Fransson \cite{Fransson20} and then further expanded in an extended series of papers (e.g., \cite{fransson2023chiral,fransson2022charge,fransson2020vibrational,fransson2023vibrationally,fransson2024current}). According to this approach, the CISS effect arises due to SOC, but is amplified by electron correlations, either by electron-electron interactions or now more commonly by interactions between electrons and chiral phonons. The starting point for the latter is the single particle SO interaction term, 
\beq
\cH_{SO}=\frac{\xi}{2}\left(\nabla V(\mathrm{\bf r}) \times \mathrm{\bf  p} \right)\cdot \bm{\si}~~,
\eeq where {\boldmath$\sigma$} is the spin operator, $\mathrm{\bf p}$ is the momentum, $\nabla V(\mathrm{\bf r})$  is the local electric field, and $\xi=\hbar/4m_e^2c^2$. One can now expand the potential around ${\bf r}_m$, the equilibrium positions of the nuclei of the molecule, which yields, to first order in the atomic position deviations ${\bf Q}_m$, a term that couples the vibrations with SOC, of the form

\beq 
\cH_1=\sum_m \frac{\xi}{2} \left\{ \nabla [{\bf Q}_m\cdot \nabla V({\mathrm{\bf r}}-{\mathrm{\bf r}^{(0))}_m}]\times \mathrm{\bf p} \right\}\cdot {\bf \si}~~. 
\eeq This expression can be further simplified by second quantization of the displacement operators and projected to a tight binding model, where it forms a spin-dependent hopping between next-nearest neighbors mediated by phonon operators (see Eq.~13 in [\cite{Fransson20}] ). The appearance of next-nearest neighbor hopping can intuitively be understood as a term that accounts for the helical curvature of the molecular structure; nearest-neighbor hopping can always be mapped to a linear system, and curvature effects only appear when the next-nearest neighbors do not lie on a straight line. 

Taking the electrode magnetization into account and realistic physical parameters, this electron-phonon CISS model demonstrates CISS polarization which can reach experimentally measured values, and qualitatively reproduces various experimental findings, including temperature and magnetic field effects \cite{das2022temperature,das2024insights}, local magnetization \cite{Zhu2024magnetic} (which was in fact {\it predicted} by the theory \cite{fransson2021charge}), and spin-dependent transport in donor-bridge-acceptor molecules \cite{chiesa2024many}. The qualitative agreement of the theory with many experimental findings has made this approach useful to interpret many recent experiments (e.g. \cite{das2024insights, sang2021temperature}).

\subsection{Central criticism - the lack of quantitative theory }
Besides the points already mentioned in the previous section, there is one central criticism which is in fact common to all CISS theories (the only exception being the spinterface theory, the topic of this review). To date, {\sl no theory has provided quantitative agreement with raw data}, such as e.g., I-V curves for different polarizations of the FM electrode. This is in spite of over a decade of transport experiments and corresponding theoretical studies. Even the electron-chiral phonon theory, a leading candidate explanation for the CISS effect, only provides qualitative agreement with experiments (see, e.g., \cite{das2024insights, das2022temperature}). The lack of such quantitative agreement is naturally accompanied by a lack of quantitative predictions. This is a severe shortcoming of all theories of CISS that the spinterface model attempts to address. In fact, we are unaware of any{\it  quantitative} predictions made by any theory other than the spinterface theory. 

Of course, the lack of quantitative agreement between theory and experiment does not rule out a theory as a viable explanation for the CISS effect. However, a higher level of confidence in a theory naturally accompanies quantitative agreement. This is thus a call for the theoretical CISS community to develop and apply theory to reproducing real raw experimental data, and for the experimental CISS community to make the raw data available in publications and to theorists.  


\section{The spinterface model } \label{The spinterface model}
\subsection{General considerations for a theory of the CISS effect} \label{General considerations}
Taking into account the different manifestations of the CISS effect - transport, photoemission, stabilization of magnetization etc. - we ask: What is the essence of the CISS effect? The answer we suggest here is that the CISS effect is the emergence of a preferred spatial direction of magnetization in the presence of chiral molecules (or, more specifically, an interface of chiral molecules and an electron reservoir). In transport experiments, this direction seems to be the direction of current flow. In photoemission and in magnetic stabilization CISS experiments, the direction is not known a priori, but is typically presumed to be perpendicular or close to perpendicular to the metal surface for the helical molecular constructs commonly used. 

It is important to recognize that inherent to CISS is not just the emergence of {\sl some} preferred direction (as in, e.g. ferromagnets); rather, the direction is provided by the geometry of the system. This point was stressed by Fransson \cite{fransson2023chiral_IJC}, who showed that in transport CISS experiments, a spin-dependent transmission is not enough to explain the CISS effect, because the direction of "up" and "down" spins is arbitrary. Only if these directions align with the preferred direction of the system {\sl and} with the magnetization directions of the FM electrode can the CISS effect be measured. 

A natural "suspect" that determines this direction is the chiral axis of the molecule. Thus, a viable theory for the CISS effect should not only demonstrate the CISS effect (preferably all of its manifestations, i.e., transport, photoemission, and appearance of magnetic moments) and demonstrate the role of chirality (i.e., the reversal of the CISS signal upon flipping of chirality), but also explain whether and how the chiral axis of the molecule defines the preferred direction. Yet, we note that this is still a conjecture (which is met by the spinterface theory, Sec.~\ref{The spinterface model}), and perhaps some other explanation can arise that does not need the chiral axis as the director axis.     

The emerging understanding of the CISS effect \cite{fransson2025chiral} is that a descriptive theory of the CISS effect must include three components: {\it i)} Molecular chirality. Chirality is necessary to break spin-degeneracy, and in the absence of chirality, the Hamiltonian describing the system is spin-symmetric, and the degeneracy cannot be broken. {\it ii)} Interactions. These are required for the enhancement of the breaking of spin-degeneracy to a measurable effect, since SOC in the molecule on its own is evidently insufficient. {\it iii)} Dissipation. This is required to stabilize the preferred spin-direction in the direction of the chiral axis. Stated differently, the appearance of a stable magnetic moment at a chiral molecule/metal interface implies broken time-reversal symmetry; if the system is dissipationless and perfectly coherent, there is no breaking of time-reversal symmetry, and a magnetic state cannot be stable. \rs{More generally, we conjecture that dissipation is crucial for the CISS effect; conceptually, it breaks time-reversal symmetry, and mathematically it allows for a stable steady-state to develop by stabilizing the feedback.} There are many plausible sources of dissipation, including electrons in the electron reservoir, phonons (either in the metal or in the molecule), nuclear magnetic degrees of freedom, etc. Quantitatively successful models of CISS must include a source of dissipation, but may assume different origins and forms for dissipation.   


\subsection{The spinterface approach}
The spinterface approach (or ``spinterface model,'' we use the two terms here interchangeably) is based on one central assumption: The presence of fluctuating magnetic moments at the interface between the electrode and the molecule, which, on average and in the absence of a chiral molecule, have no preferred direction. According to the spinterface approach, in all its manifestations, the CISS effect amounts to stabilizing these surface magnetic moments to a specific direction, defined in this model by the molecular chirality and orientation. 

How does this stabilization take place? There are three ingredients to the process. The first is the motion of electrons within the chiral molecule, which can be either due to external stimuli, e.g., a voltage drop, or spontaneous, as discussed in the following sections. Regardless of the cause for electron motion, as electrons pass in or out of the chiral molecule, they generate a ``solenoid field'' - a magnetic field that has a component parallel to the direction of the average electron current, and similar to the magnetic field generated in a classical solenoid. As discussed in Sec.~\ref{What is the theoretical difficulty?}, this field is tiny and cannot be solely responsible for the CISS effect. Rather, we hypothesize that its only role is to break the directional symmetry, giving rise to a preferred direction. 

The second ingredient, which enhances the directional symmetry-breaking effect of the solenoid field, is magnetic exchange interactions between the surface magnetic moments and the electrons in the molecule. These interactions act across the electrode-molecule interface - hence the name spinterface approach - and amplify the effect of the solenoid field. They may depend on the nature of the surface magnetic moments, and can be either SOC or electron exchange interactions, resulting in both cases in an effective spin-dependent shift in the orbital energies of the molecule, or, equivalently, a spin-dependent shift in the chemical potential. The third ingredient is a dissipative term acting on the surface magnetic moments. In the absence of a magneto-dissipative term, any interactions that the moments experience can only lead to coherent rotations around a given axis. These rotations are suppressed by dissipation, leading finally to the stabilization of the surface moments in the direction determined by the solenoid field. 

\rs{It is imperative to highlight that the spinterface model, therefore, does not contradict supramolecular length dependence: the interface provides the necessary symmetry breaking, while the transport mechanism through the bulk chiral medium governs how the effect scales with molecular length or supramolecular organization. Consistent with this, a single set of spinterface parameters quantitatively fits transport data across systems of differing molecular thicknesses \cite{dubi2022spinterface}.}

In the following subsections, we describe two possible spinterface mechanisms, which we dub the ``dynamical'' and the ``displacement'' spinterface mechanisms. The difference describes the origin of the electron motion leading to the stabilization of the surface magnetization: Either current passing through the junction due to an applied voltage or displacement currents due to level misalignment between the molecular orbitals and the electrode chemical potential.

\subsubsection{The dynamical spinterface model}
The dynamical model (introduced in \cite{alwan2021spinterface}) aims at explaining the transport CISS effect in molecular junctions. It assumes that at equilibrium,  the surface magnetic moments at the electrode-molecule interface are strongly fluctuating with nothing to stabilize them (in contrast to the displacement spinterface model, see next section). 

Once a voltage bias is applied across the junction, current begins to flow, and as it flows through the chiral molecule, it generates a solenoid magnetic field. This field is rather small, but it is enough (\rs{orders of magnitude higher than the earth's magnetic field, see Sec. \ref{sec:earth_mag_fld}}) to break the symmetry and generate a tiny tilt in the surface magnetic moments. These moments interact magnetically with the electrons, or - equivalently - electrons scatter off the magnetic moments as they cross the interface. Now, depending on the chirality of the molecule, the solenoid field will tilt the surface moments either parallel (i.e., spin-up) or anti-parallel (spin down) to the direction of the current flow. Consequently, one spin-species will have slightly higher conductance than the other. 

The result of this is that there will be a small spin imbalance in the molecule near the electrode-molecule interface. This spin imbalance, interacting through SO or exchange interaction with the surface moments, acts as an additional magnetic field, which tilts the surface magnetic moments further, in analogy to an applied spin-torque. The additional tilt of the surface magnetization generates an even larger spin--imbalance, and so on and so-forth. This feedback loop amplification mechanism - a core aspect that resolves the problems with other models that rely e.g. on SOC in molecules - reaches a steady-state in which the magnetization is tilted, generating a different barrier for electrons entering or leaving the molecule with different spins, thus generating a measurable CISS effect. 

Before briefly describing the mathematical form of this mechanism, we comment here on the symmetry of the effect. As pointed out in Sec.~\ref{What is the theoretical difficulty?} (Fig.~\ref{fig:IV-toymodel}), to capture the transport CISS effect, the effective barrier felt by the different spin-species must be symmetric with respect to the bias voltage, while a barrier which directly depends on the solenoid field is linear in voltage (i.e., demonstrating the MChA effect). To understand why the dynamical spinterface model 
has the correct symmetry, we note that the spinterface is on one side of the molecule only. Without loss of generality, let us assume that for positive bias, electrons move from the molecule towards the spinterface, and that the chirality of the molecule is such that spin-up electrons are aligned with the tilt of the magnetization, thus scattering more efficiently through the molecule and favoring spin-down electrons in the molecule. When the current is reversed, the direction of the solenoid field is reversed, and thus the tilt of the magnetization is such that now spin-down electrons scatter less. However, they now flow from the electrode into the molecule, so again the result is a spin imbalance favoring spin-down electrons in the molecule. This implies that regardless of the direction of current, the spin-torque exerted on the surface magnetization by the spin imbalance will depend only on the chirality of the molecule and not on the direction of current, which is the symmetry expected in the transport CISS effect. 


Casting the dynamical spinterface approach into mathematical form is conceptually rather simple. The first requirement is a model of the specific CISS system at hand, be it a single-molecule junction, a multi-molecule junction, a chiral layer, etc. By "model", we mean that one should have a formulation of the current-voltage characteristics, $I_0(V)$ (we call this the "bare I(V)"), and the electron population (or density) in the system, $n(V)$, as a function of the external voltage $V$, and other parameters of the system. Importantly, the current and density should depend in some way on the scattering through the interface. A natural form for such a dependence is to represent the interface as subject to tunneling through a barrier by using simple Arrhenius formalism \cite{alwan2023temperature} or a Simmons model \cite{vilan2007analyzing,vilan2013rethinking}), using a simple classical tunneling description\cite{dubi2022spinterface} or using a Lorentzian transmission within the Landauer formalism (Eq.~\ref{eq:toymodel})\cite{alwan2024role}). 

In all these examples, the electron transport across the spinterface depends on the barrier energy, or, equivalently, on the electrode chemical potential, $\mu$. The dynamical spinterface model implies that the interaction of the surface magnetic moments with the electrons changes the chemical potential or, equivalently, the molecular orbital energy in a spin-dependent manner, such that the energy splitting between the two spin species (denoted by \(\sigma = \pm 1\)) is described by
\beq
   \mu_\sigma = \mu_0 + \sigma \alpha_A \cos(\theta_M)
    \label{eq:zeeman_splitting}
\eeq
where \(\mu_\sigma\) is the chemical potential for spin species \(\sigma\), \(\mu_0\) is the reference chemical potential, \(\alpha_A\) is the interaction strength between the surface moments and electron spins (in eV), and \(\theta_M\) is the tilt angle of the orbital magnetic moment relative to the molecular axis. The tilt angle \(\theta_M\) is in turn determined by the torque applied on the surface magnetization from the solenoid field and the spin imbalance, which generates an effective magnetic field, 
\beq
    m_0 B_{\text{eff}} = \alpha_0 J + \alpha_1 \Delta s
    \label{eq:effective_field}
\eeq
where \(m_0\) is the Bohr magneton, \(\alpha_0\) represents the strength of the solenoid field, \(\alpha_1\) is the coefficient for the spin-transfer torque, and \(\Delta s = n_\uparrow - n_\downarrow\) is the spin imbalance, with \(n_\uparrow\) and \(n_\downarrow\) representing the populations of spin-up and spin-down electrons, respectively.

The self-consistent nature of the calculation is now visible: The spin-dependent chemical potential $\mu_\sigma$ depends on the magnetization tilt angle $\theta_M$, which depends on the current and density in the molecule, which in turn depend on the spin-dependent chemical potential and on the voltage bias across the junction. Thus, a self-consistent calculation is required in order to evaluate the steady-state currents, densities, and tilt angles. The calculation will naturally depend on the model details, but a general approximation can be made by assuming that the tilt-angle is subjected to thermal fluctuations, leading to a simplified formula \cite{alwan2021spinterface} 
\begin{equation}
\cos(\theta_M)=\frac{\bessel_1(\xi)}{\bessel_0(\xi)}, ~\xi=\frac{\alpha_0 J[\theta_M]+\alpha_1 \Delta s[\theta_M] }{k_B T}~~,
\label{eq:Bessel}\end{equation}
where $\bessel_0, \bessel_1$ are the Bessel functions of the first kind. Eq.~\ref{eq:Bessel}, together with a specific model for the current and density through the junction, provides a fully closed mathematical system to evaluate the currents in a CISS transport experiment. 

\rs{It is important to note that within the dynamical spinterface model, $\alpha_0$ and $\alpha_1$ are the only fit parameters that determine the strength of the measured CISS effect. Typically, values of $\alpha_0 \sim 0.1-1$ eV and $\alpha_1 \sim 1-100$ meV are obtained from fits to the experiment (see Section \ref{Th_n_Exp}), which are realistic typical values for experimental systems.}
\subsubsection{The displacement spinterface model}
The \rs{dynamical spinterface model} describes very well the CISS effect in transport experiments (see Sec.\ref{Th_n_Exp}). However, a key feature of the model is that in the absence of current through the system (i.e. if the bias voltage is zero), there is no effect because the stabilization of magnetic moments dynamically depends on the electron current and density. 

This raises a natural question: if bias voltage is required to generate the CISS effect, what is the mechanism for the CISS effect observed in photoemission experiments? Can the spinterface mechanism also be relevant for the photoemission CISS effect? 

Recently, some of us addressed this question, demonstrating that indeed the spinterface model can be used to understand the photoemission CISS effect \cite{monti2024surface}. The idea is that the spinterface mechanism - i.e. the stabilization of surface moments due to interaction with electron spins in the molecule, with the solenoid current breaking the initial symmetry - is exactly the same, and the only essential difference is that what drives the electron current is not bias voltage, but rather "displacement currents", i.e. the spontaneous currents to or from the molecule to the electrode upon initial contact. 

The most important feature of this so-called "displacement spinterface model" is that it occurs spontaneously as chiral molecules are attached to the metallic surface, and the result is a stable magnetic moment at the interface between the molecules and the electrode. We showed \cite{monti2024surface} that this can lead not only to a photoemission CISS effect consistent with experiments, but also to quantitative predictions. 

The key result of the displacement spinterface model can be laid out mathematically with great simplicity. It amounts to a constant spin-dependent shift in the chemical potential of the metal electrode, or equivalently of the molecular energy level, in the form of 
\beq
\varepsilon_\si=\varepsilon_0 + \si \chi ~~,\label{energy shift}\eeq
where $\varepsilon_0$ is the bare molecular frontier orbital energy, $\si$ is the electron spin and $\chi$ is the energy shift associated with the magnetic interactions between the surface magnetization and electron spins in the molecule. Importantly, a constant energy shift is consistent with the symmetry of the CISS effect (i.e. even in voltage), and displays the correct phenomenology, as we will demonstrate in what follows.

\subsection{The spinterface model - Theory and Experiment}\label{Th_n_Exp}

As pointed out earlier, the spinterface model is to date the only theoretical approach to the CISS effect that provides fits and reproduces experimental data. Here we review some examples of these fits. Importantly, we note that the spinterface model can be used to fit a very wide variety of experimental systems at different scales and structures, from small single molecules \cite{yang2023}, through large bio-molecules \cite{dubi2022spinterface},  PANI macro-structures \cite{sarkar2022temperature}, intercalated chiral layers \cite{alwan2024role}, and more. However, it is essential to note that the spinterface theory relies on some system-dependent parameters. These must be physically reasonable in magnitude and can be extracted from experiments by fitting (vide infra). The spinterface approach does not provide a framework for calculating these parameters for different systems from first-principles calculations (see also the discussion in Sec.~\ref{critical}). 

The procedure for fitting experimental data to the spinterface theory starts with fitting the bare I-V curves to a specific model, which changes depending on the system under study. For instance, for single-molecule junctions, a single-level quantum Landauer transport model is applicable, while for biomolecules, a multi-level hopping model, quite expectedly, shows much better agreement with the bare I-V curves. Once the I-V curve is modeled, the spinterface theory is used to evaluate the total current for the two magnetization directions of the FM electrodes, $J_s(V)$. 

\begin{figure}
\begin{center}
\includegraphics[width=10cm]{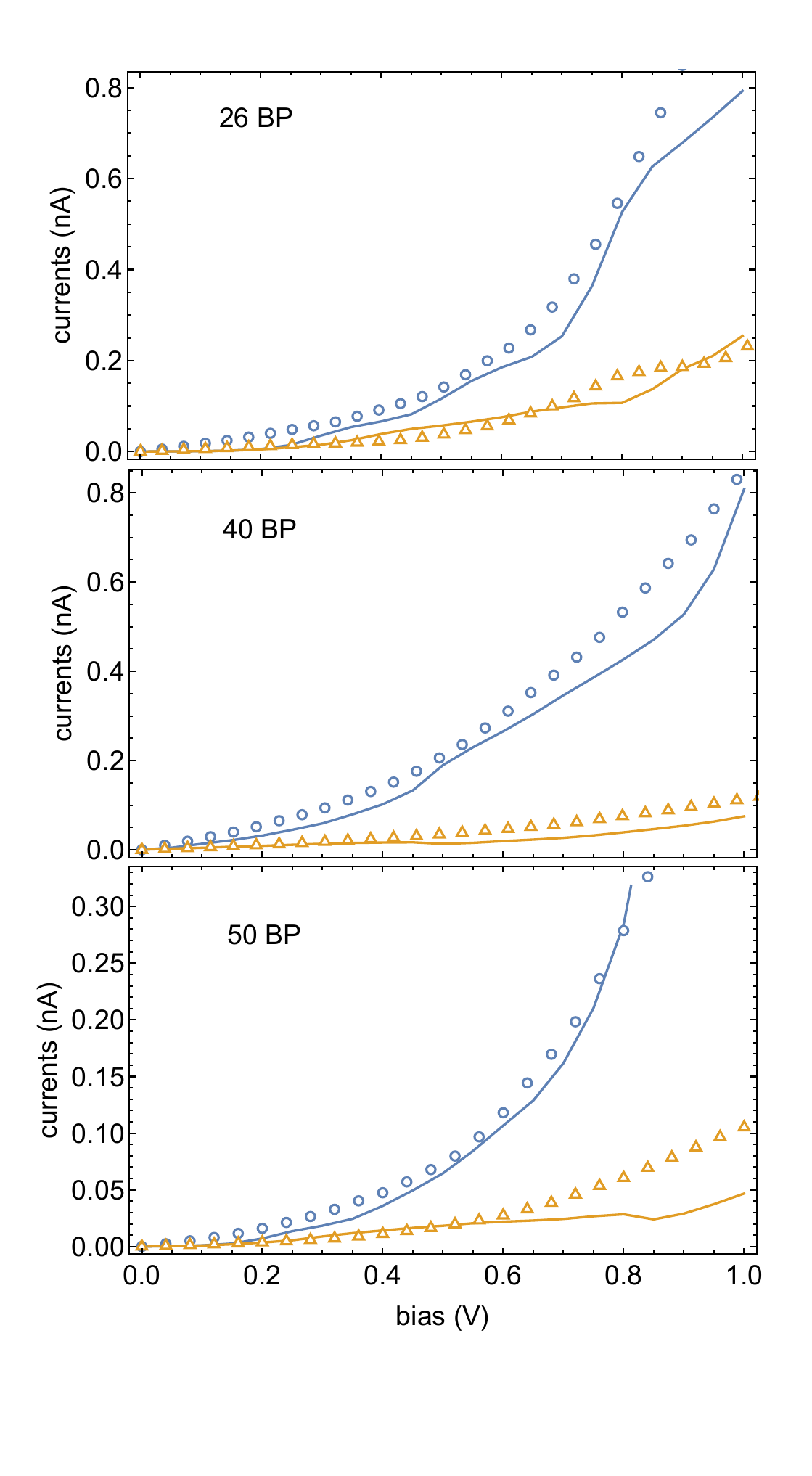}
\caption{
I-V curves for two directions of FM electrode magnetization, parallel (blue) and anti-parallel (orange) to the direction of current flow. Open symbols are the experimental data {(adapted with permission from Xie, Z.; Markus, T. Z.; Cohen, S. R.; Vager, Z.; Gutierrez, R.; Naaman, R. Nano Lett. 2011, \textbf{11}, 4652–4655. Copyright 2011 American Chemical Society; Ref.~\onlinecite{Xie11})} and solid lines are the theoretical fit to the spinterface model {(adapted with permission under a Creative Commons CC BY 4.0 License from Dubi, Y. Spinterface chirality-induced spin selectivity effect in bio-molecules. Chem. Sci. 2022, \textbf{13}, 10878–10883. Copyright 2022 The Author(s) – Published by the Royal Society of Chemistry; Ref.~\onlinecite{dubi2022spinterface})}.
}
\label{fig_ChemSci2022}
\end{center}
\end{figure}

The first example (Fig.~\ref{fig_ChemSci2022}, taken with permission from Ref.~\onlinecite{dubi2022spinterface}) pertains to data measured in the Na'aman group \cite{Xie11}, where the CISS effect was measured in DNA oligomers with different numbers of base pairs (open symbols in Fig.~\ref{fig_ChemSci2022}). The solid line represents the theoretical calculation of the raw data. The model assumes classical multi-level hopping as the transport mechanism \cite{dubi2022spinterface}. Importantly, the CISS parameters, namely the electrode SO coupling and the exchange interaction term, neither of which are affected by the length of the molecules, are reasonable and constant for all the calculations, and only the molecular parameters extracted from the bare-IV curves are changed as a function of the number of base-pairs. The resulting agreement between theory and experiment is quite striking, and Ref.~\onlinecite{dubi2022spinterface} contains additional fits to more recent data \cite{mishra2020length}.  

\begin{figure}
\begin{center}
\includegraphics[width=10.6cm]{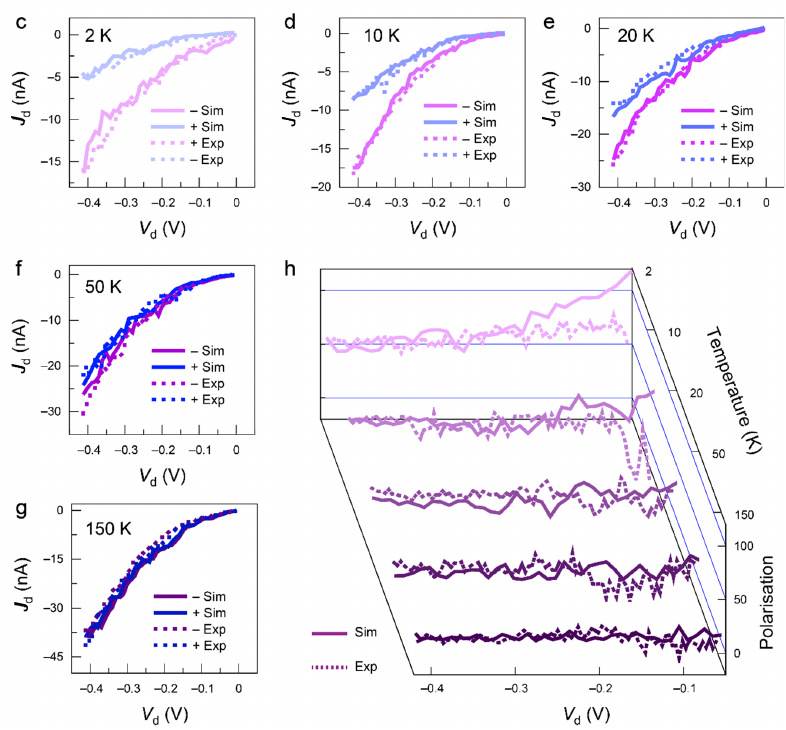}
\caption{I-V curves for two directions of FM electrode magnetization, parallel (purple) and anti-parallel (blue) to the direction of current flow. Dots are the experimental data and the solid lines are the theoretical fit to the spinterface model {(adapted with permission from Yang, C.; Li, Y.; Zhou, S.; Guo, Y.; Jia, C.; Liu, Z.; Houk, K. N.; Dubi, Y.; Guo, X. Nat. Chem. 2023, \textbf{15}, 1–8. Copyright 2023 Springer Nature.; Ref.~\onlinecite{yang2023}).}
}
\label{fig_NatChem2023}
\end{center}
\end{figure}

A second example comes from Ref.~\onlinecite{yang2023}, where a single maleimide molecule is placed in a molecular junction consisting of one FM and one paramagnetic electrode, treated by a Michael addition to form a chiral moiety, and constantly monitored by current measurements \cite{yang2023}. Figure \ref{fig_NatChem2023}  shows the I-V curves for the two directions of the FM magnetization, parallel (purple) and anti-parallel (blue) to the direction of current flow. Dots are the measured data, and the solid lines are the theoretical calculations (where the bare I-V curves are reproduced from interpolating measurements \cite{yang2023}). 

In the figure, measurements are shown for temperatures ranging from 2K up to 150 K. Importantly, the fit to extract the CISS parameters is only performed on the 2K data. In other words, the remarkable agreement between theory and experiment for all temperatures 10-150K is achieved without any additional fitting parameters. The agreement between theory and experiment can also be seen in the bottom-right part of the figure, where the polarization as a function of voltage is plotted for different temperatures (again,  no additional fit parameters are used above 10 K). The temperature dependence of the CISS effect 
is clearly seen here, showing that increasing temperature reduces the CISS effect (see discussion in Sec.~\ref{critical}). 

\begin{figure}
\begin{center}
\includegraphics[width=10.6cm]{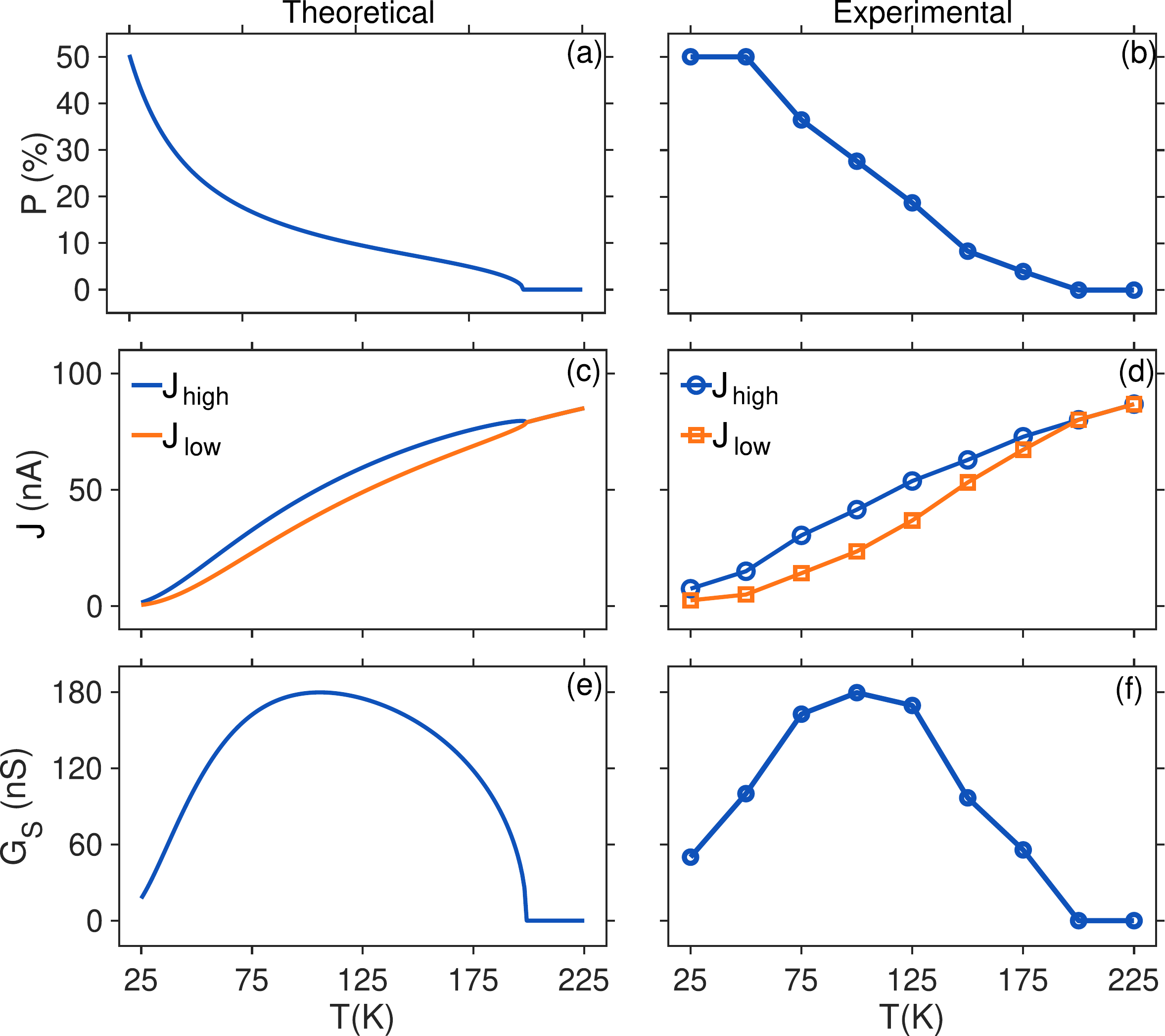}
\caption{Experimental data (right panels) and theoretical calculation (left panel) for the CISS polarization, currents and "spin-conductance" of a layer of chiral molecules intercalated between layers of two-dimensional material. The results show excellent quantitative agreement between theory and experiment, and support the claim that the CISS effect is reduced with increasing temperature. {(reprinted with permission under a Creative Commons CC BY 4.0 License from Alwan, S.; Sarkar, S.; Sharoni, A.; Dubi, Y. J. Chem. Phys. \textbf{159}, 014106 (2023). Copyright 2023 The Author(s), published by AIP Publishing; Ref.~\onlinecite{alwan2023temperature}
} ) 
}
\label{fig_JPCm2023}
\end{center}
\end{figure}

A third example comes from Ref.~\onlinecite{alwan2023temperature}, pertaining to experiments by Qian et al.\cite{qian2022chiral}, who report intercalating a layer of chiral molecules between the layers of layered two-dimensional atomic crystals, and measuring the current through the system at constant voltage and hence conductance for two magnetization directions of an FM electrode. From this, the polarization and a term defined as "spin-conductance", $G_S$, were reproduced. In Fig.~\ref{fig_JPCm2023} (reproduced with permission from Ref.~\onlinecite{alwan2023temperature}) we show side by side the experimental data (right panels) and theoretical calculations (left panels), demonstrating once again excellent agreement between theory and experiment. As described in detail in Ref.~\onlinecite{alwan2023temperature}, our results demonstrate that the interpretation of the authors that the CISS effect increases with increasing temperature is in fact erroneous, and is a result of misuse of the Julli\'ere formula. In fact, the data of Ref.~\onlinecite{qian2022chiral} supports the claims that the CISS effect is reduced with increasing temperature (see discussion in Sec.~\ref{critical}).

\begin{figure}
\begin{center}
\includegraphics[width=10.6cm]{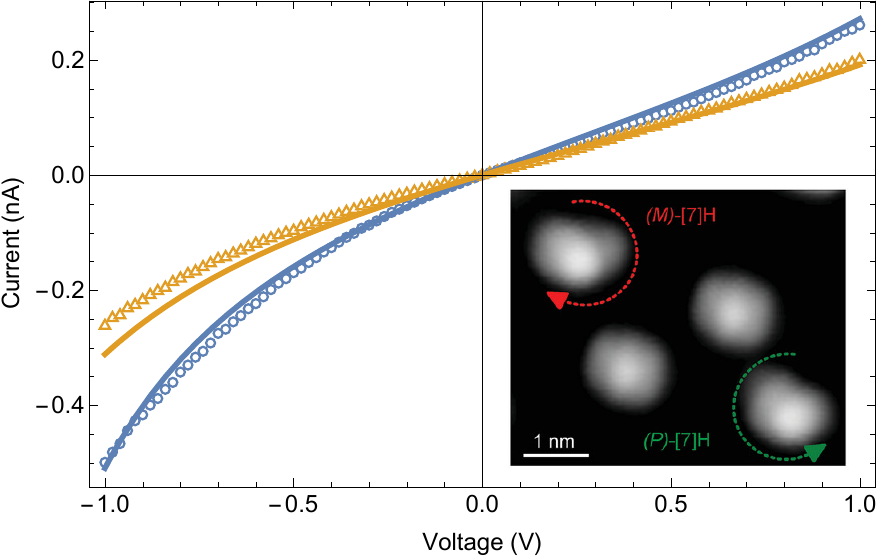}
\caption{Experimental data (symbols) and theoretical calculation (solid lines) for the I-V measurements of Helicenes on Co nano-islands (data courtesy of the authors of Ref.~\onlinecite{safari2023spin}), for two helical directions. The theory used for the fit is the dynamical spinterface model, which considers the generation of a constant surface magnetization (independent of voltage). Inset: STM image of the chiral molecules on the surface, taken with permission from Ref.~\onlinecite{safari2023spin}. {Both the data and the STM image are adapted with permission under a Creative Commons CC BY 4.0 License from Safari, M. R.; Matthes, F.; Schneider, C. M.; Ernst, K.-H.; Bürgler, D. E. \textit{Small} 2023, \textbf{19}, 2308233. Copyright 2023 The Author(s), published by Wiley-VCH GmbH.} ) 
}
\label{Buergler_fit}
\end{center}
\end{figure}

In these three examples and other examples in \onlinecite{alwan2024role}, the transport spinterface model (sec.~\ref{The spinterface model}) was used to fit the experiments. Next we show that the dynamical spinterface model \cite{monti2024surface}, which implies that there is a constant magnetization at the surface independent of voltage, can also be used to fit relevant experimental data, including from transport experiments. To show this, we refer to a recent paper by Safari et. al., who measured the CISS effect of helicenes placed on Co nano-islands \cite{safari2023spin} (see inset in Fig.~\ref{Buergler_fit}, taken with permission from Ref.~\onlinecite{safari2023spin}). To fit the data, we assume that transport through the chiral molecule is characterized by the low-temperature limit for transport through a single level \cite{Book:Cuevas_Scheer10}, with the form 

\begin{equation}
J(V; \varepsilon_0)=\frac{1}{\Gamma}\left( tan^{-1}\left(\frac{\varepsilon_0+\alpha V +V/2)}{\Gamma} \right) - tan^{-1}\left(\frac{\varepsilon_0+\alpha V -V/2)}{\Gamma} \right) \right) ~.\label{J(V)_lowT}\end{equation}
Here, $\Gamma$ is the level broadening, found from the data to be $0.015$eV, which is a small value, common to molecules which are physisorbed on a surface,  $\varepsilon_0$ is the energy of the frontier molecular orbital $\varepsilon_0=1.03$eV, and $\alpha=-0.107$ is an asymmetry parameter, reflecting the distribution of potential along the molecule leading to an asymmetric I-V curves (see, e.g.,\onlinecite{xu2015negative}). 

According to the dynamical spinterface model \cite{monti2024surface}, when a chiral molecule is placed on a surface, electrons flow between the molecule and the electrode exerts a spin-torque by exchange or SO interaction on the surface moments of the electrode, stabilizing them in the chiral direction of the molecule. The effect on transport is thus a spin-dependent shift in the orbital energy, of the form $\varepsilon_0 \rightarrow \varepsilon_0+\sigma \chi$, where $\chi$ is the energy shift due to the local magnetic field induced by the stable magnetic moments or, equivalently, the spin-exchange interaction at the interface, and $\si$ is the electron spin. The resulting current for the two chiral directions $s=\pm 1$ pointing approximately parallel and anti-parallel to the surface normal, and taking the finite polarization of Co ($P_{co}\sim 0.15$, or $15\%$) into account, is 
\beq
J_s(V)=(1-P_{Co} )J(V; \varepsilon_0+s \chi)+P_{Co}J(V; \varepsilon_0-s \chi) ~~. \label{J(V)_fit_dynamicalCISSmodel}
\eeq

The two colors in Fig.~\ref{Buergler_fit} correspond to the currents,$J_s(V)$, for the two chiral directions. By fitting the data, we find $\chi\sim0.1$ eV. 

We note that in these experiments, the molecules are placed directly on an FM. One can thus ask what the difference is between an FM and a regular metal in generating stable surface magnetization? To examine this question, we calculated the dynamics of surface magnetization stabilization (as in Ref.~\onlinecite{monti2024surface}), now including spin-dependent transfer rate constants from the molecule into the electrode, reflecting the difference in the density of states for the two spin species in the Co nano-island. We find that besides small quantitative changes in the dynamics, there is no difference in the final configuration of the magnetic moments. Put simply, we show that the surface magnetic moment stabilization, driven by the spinterface mechanism, works in metallic ferromagnets as well as paramagnets. \rs{This is further corroborated by the recently observed CISS--MR with non--magnetic contact\cite{hossain2025observation}.}   

\begin{figure}
\begin{center}
\includegraphics[width=10.6cm]{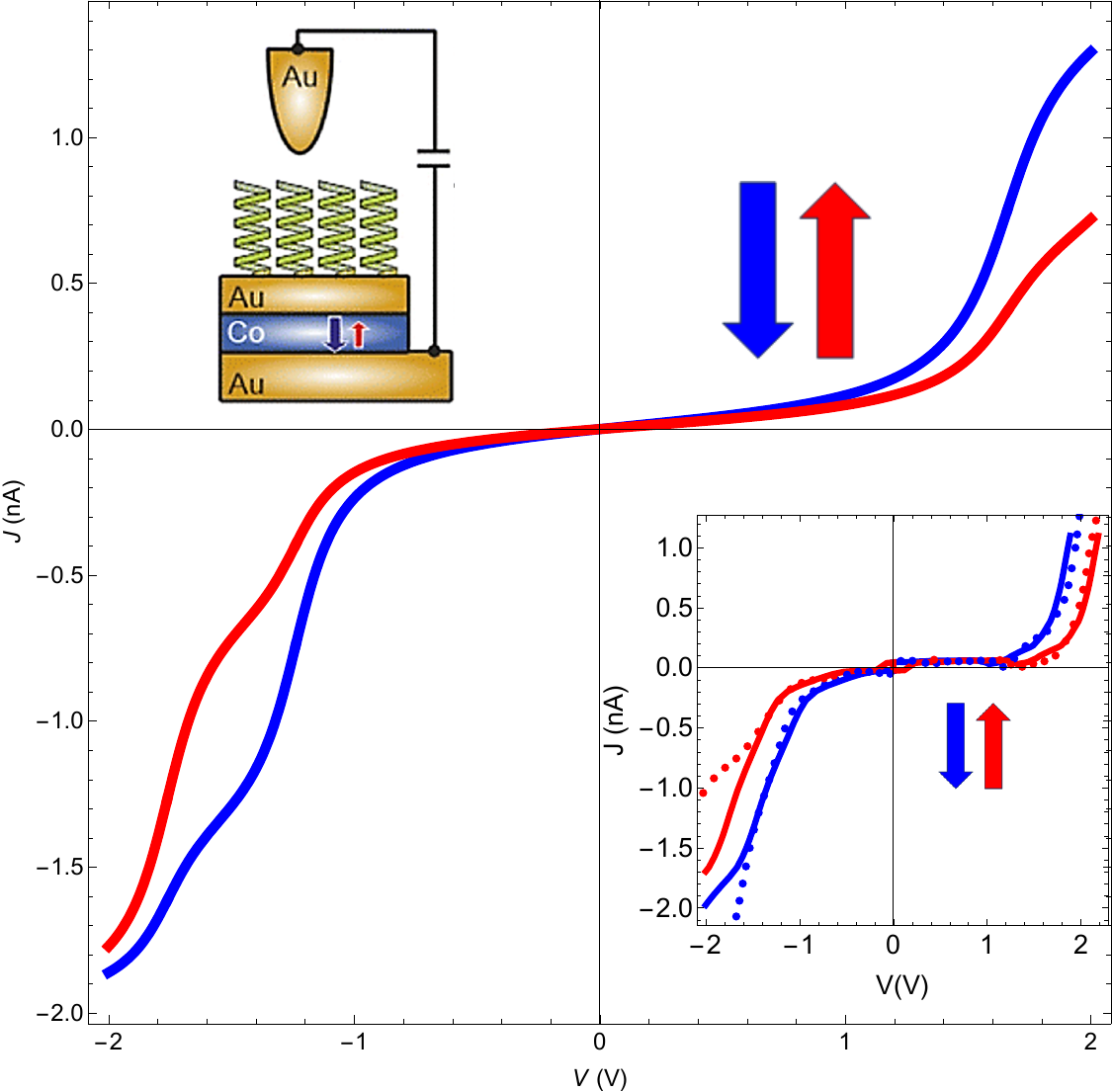}
\caption{ I-V curves for two magnetization directions using the dynamical spinterface model Eqs.~(\ref{J(V)_lowT}-\ref{J(V)_fit_dynamicalCISSmodel}), with parameters extracted from the data of Ref.~\onlinecite{nguyen2024mechanism}. Top-left inset: experimental setup. Top-right inset:  Bottom-right inset: correspondence between theory (solid lines) and data (dots, experimental data courtesy of the authors of Ref.~\onlinecite{nguyen2024mechanism}), when the I-V curve is directly extracted from the measurement through interpolation. {Experimental data adapted with permission under a Creative Commons CC BY 3.0 License from Nguyen, T. N. H.; Salvan, G.; Hellwig, O.; Paltiel, Y.; Baczewski, L. T.; Tegenkamp, C. Chem. Sci. 2024, \textbf{15}, 14905. Copyright 2024 The Author(s) – Published by the Royal Society of Chemistry.}) 
}
\label{Tegenkamp_fit}
\end{center}
\end{figure}

In a final example, we turn to recent data by Nguyen et al. \cite{nguyen2024mechanism}, who measured transport through helical polyalanine molecules in contact with an Au AFM tip and an Au/Co/Au substrate. The latter allows polarization control in the Co layer, parallel/anti-parallel to the direction of current flow, see top-right inset in Fig.~\ref{Tegenkamp_fit}. The experimental results in the form of I-V for the two magnetization directions are shown at the top-left inset of Fig.~\ref{Tegenkamp_fit}. To fit these data, we again consider the model of Eqs.~(\ref{J(V)_lowT}-\ref{J(V)_fit_dynamicalCISSmodel}). Fitting the data, we find the fit parameters $\Gamma\sim 0.08$ eV, $\varepsilon_0\sim 0.85$  eV, and $\alpha\sim 0.072$. \rs{The value of level broadening $\Gamma$ is consistent with reported molecule–electrode couplings, which range from sub-meV to several tens of meV in experiments \cite{gorenskaia_methods_2024, sendler_light-induced_2015, yelin_conductance_2016, garrigues2016single, baldea2024dichotomy}.} A spin-exchange term of $\chi\sim 0.15-0.2$ eV is found to faithfully represent the experimental data. In the bottom-right inset of Fig.~\ref{Tegenkamp_fit}, we show the correspondence between theory (solid lines) and data (dots), when the I-V curve is directly extracted from the measurement through interpolation. 

\begin{figure}
\begin{center}
\includegraphics[width=10.6cm]{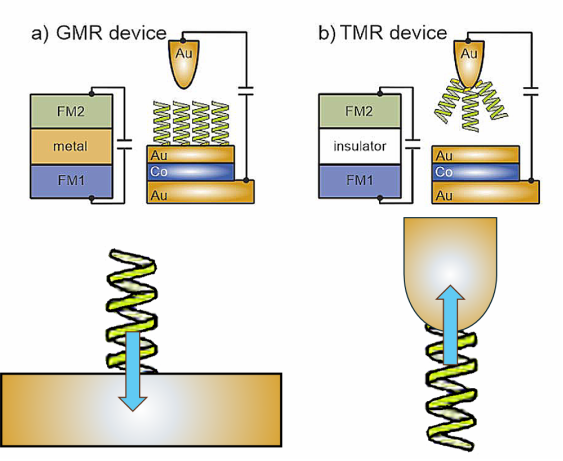}
\caption{The dynamical spinterface model \cite{monti2024surface} provides a simple explanation for the observation of Nguyen et al. \cite{nguyen2024mechanism}, that in GMR vs TMR devices the role of majority vs minority spin is reversed and the CISS polarization flips, as if the chirality of the molecules was reversed (see text for details). {Schematics of the GMR and TMR device are adapted with permission under a Creative Commons CC BY 3.0 License from Nguyen, T. N. H.; Salvan, G.; Hellwig, O.; Paltiel, Y.; Baczewski, L. T.; Tegenkamp, C. Chem. Sci. 2024, \textbf{15}, 14905. Copyright 2024 The Author(s) – Published by the Royal Society of Chemistry.}) 
}
\label{Tegenkamp_explain}
\end{center}
\end{figure}

While the quantitative agreement between the data of Ref.~\onlinecite{nguyen2024mechanism},
and the spinterface theory is quite clear, there is yet another experimental finding by Nguyen et. al., which can be easily understood within the spinterface model, yet is challenging to account for with other explanations of the CISS effect. In their experiments, Nguyen et al. measure the transport CISS effect with two experimental setups. In the first, which they call "GMR device", chiral molecules are placed on the Au-Co-Au (i.e., FM) layer, and are approached by the Au-AFM tip for transport measurements. In the second setup, which they call "TMR device", chiral molecules cover the Au-AFM tip, and then the decorated AFM tip approaches the FM surface for transport measurements (see Fig.~\ref{Tegenkamp_explain}, taken with permission from Ref.~\onlinecite{nguyen2024mechanism}). Naively, one would not expect any difference between these measurements. Still, the authors surprisingly find that when measuring the transport CISS effect for these devices, the role of majority and minority directions is reversed when comparing the GMR and TMR devices. Put simply, if in the GMR device the high-current I-V curve appears when the FM is magnetized parallel to the current flow; then in the TMR device, the high-current I-V curve will appear when the FM is magnetized anti-parallel to the direction of current flow. Note that other than during the initial approach, at the point of current measurement the GMR and TMR systems are believed to be essentially the same, and yet  they behave as if the chirality of the molecule has been reversed. This observation cannot be accounted for by any theory for the CISS effect which does not include an electrode effect. 

The spinterface theory, and specifically the dynamical spinterface model \cite{monti2024surface}, actually provides a simple explanation for this observation. We remind the reader that the theory suggests that as the molecule is placed on the metal electrode, electrons flow (say) from the molecule to the electrode, and this displacement current applies via exchange interactions or SOC a torque on the surface magnetic moments, and stabilizes them along the current. Now consider the case of GMR vs TMR devices (Fig.~\ref{Tegenkamp_explain}). Since the molecules are attached to the surface from below (GMR) and from above (TMR), and since the direction of displacement current depends only on the properties of molecule and surface and is the thus same (say from the molecule to the metal), clearly its direction is opposite for the two cases. Consequently, the surface magnetic moments will be stabilized in two opposite directions. Thus when contact is made for transport measurements, the GMR and TMR  already have stabilized surface magnetic moments pointing in opposite directions. This means that when the current is positive (e.g. from the FM to the AFM tip) in the GMR setup electrons with spin parallel to the direction of current will have lower energy because of the interaction with the localized surface moments, but in the TMR setup they will have higher energy. Put simply, the role of majority and minority spins has been reversed. 

\subsection{The spinterface model - A critical view}\label{critical}
The fact that the spinterface theory can account for numerous experimental findings and indeed reproduce experimental raw data, with realistic parameters and to unprecedented accuracy, gives confidence that it is a faithful description of the physical origin of the CISS effect. Nonetheless, the theory is not without flaws and challenges. It is important to acknowledge them openly, which we believe is the best way to make progress. Therefore, in this section we aim at listing some of the main points of criticism raised against the spinterface model, and provide our rebuttal to the critique. 

\subsubsection{Origin of surface magnetic moments}
A central assumption of the spinterface approach is that there are magnetic moments at the surface of the metal electrode, which typically have no preferred orientation, because of disorder or fluctuations. The assumption is that they are susceptible enough such that the spinterface interaction can stabilize them. A relevant question to ask is: What is the origin of these magnetic moments, and what is their nature? 

We cannot give a definite answer to this question. These moments may be related to localized surface states, localized by surface disorder or by contact with the molecule. They may be related to the spin-orbit interaction on the surface, to dangling bonds, to the hybrid nature of the surface orbitals in metals, etc. Further research is required to determine the microscopic origin of the surface states and their magnetic properties.  What {\sl can} be said with confidence is that it is widely known that in metallic nano-structures in general, and Au in particular, the surface indeed develops magnetic properties \cite{nealon2012magnetism, trudel2011unexpected, singh2013unexpected, krishna2014chemically, agrachev2017magnetic, ulloa2021magnetism, dong2020tuning, tuboltsev2013magnetism, cheng2020light, gessner2022magnetic, banerjee2009gold, donnio2017characterization}. These magnetic properties develop spontaneously or can also be triggered by illumination), and can be controlled by controlling the shape and size of the nanoparticles or by chemically coating them. 

The origin of magnetism in metallic nanoparticles is unknown, and may be an orbital effect \cite{banerjee2009gold, gomez2018orbital}, a quantum effect related to persistent Eddy currents \cite{greget2012magnetic} or a plasmonic effect \cite{hurst2018magnetic}. Research in this field is quite active, and it may be that the CISS effect is another manifestation of this phenomenon. 

\subsubsection{Magnitude of magnetic fields induced by current through chiral molecules}\label{sec:earth_mag_fld}
The spinterface mechanism assumes that the CISS measurements are enabled due to a cascade of energy scales, starting from the small energy scale related to the magnetic field generated at the interface due to the current through the chiral molecule (i.e. a "solenoid" field"), through the spinterface exchange interactions, to the SO interaction in the electrode. A natural question to ask is how can the current through a "molecular solenoid", when the current is small and the solenoid is small, affect the electronic properties at all? And, if the system is so susceptible to the magnetic field to break the directional symmetry, how is it that the earth's magnetic field ($\sim 50\mu$T) does not overcome the current-induced field? 

When considering the current-induced magnetic field in the situation described in the dynamical spinterface model, one can evaluate the effective magnetic field with the solenoid formula, $B\approx \mu_0 N_t J/L$, where $\mu_0=4\pi \times 10^{-7}$T~m/A is the permeability of free space, $N_t$ is the number of turns, $L$ the molecular length (typically $N_t/L\sim 10^9$) and $J$ is the displacement current. One can estimate the displacement current by noting that a single electron (i.e. total charge $\sim 1.6 \times 10^{-19}$C) passes through the interface at a typical timescale of $~1-10$fs, leading to a field pulse of $\sim 100$mT, orders of magnitude larger than the earth magnetic field, and probably much larger than the FM fringe field. 

\subsubsection{Origin of a "solenoid" magnetic field in non-helical molecules}
A key ingredient of the spinterface model is the generation of a "solenoid" field in the molecule, which breaks the directional symmetry between spin parallel and anti-parallel ("up" and "down") to the direction of overall current flow. While in a molecule with a {\sl helical} structure one can imagine that currents, either due to electron displacement or due to a bias voltage, can generate a magnetic field in the direction of current flow, this effect is harder to imagine in chiral yet non-helical molecules. Nonetheless, the CISS effect was measured with such molecules \cite{yang2023}, and the spinterface model was able to reproduce these data with remarkable accuracy, see Fig.~\ref{fig_NatChem2023}. 

Current work (in the spirit of Ref.~\onlinecite{bro2025substantial}) is underway, to answer exactly this question: can a magnetic field be generated in the direction of current) even in non-helical yet chiral molecules, what are the conditions required, and how is such a magnetic field  related to the molecular structure. Since the only requirement for generating some solenoid field is that some part of the current density follows a helical pathway, the generation of solenoid fields in chiral non-helical molecules seems very likely, yet this should be tested via computational methods. At a deeper level, this should be connected to the symmetries of the system \cite{theiler2025non}.  

We take this opportunity to clarify the notion of chirality in molecules in the context of electron transport. Clearly, when we say that a molecule is chiral, what we mean is that it is {\sl structurally} chiral, i.e. that the atoms in the molecule are arranged in a chiral arrangement, i.e. the molecule has two mirror symmetric forms which cannot be overlayed by rotations only. However, this does not necessarily mean that electron currents move along chiral trajectories. An example for that is transport through helicenes, which are clearly structurally chiral. However, when the helicene molecules are squeezed in the vertical direction, such that the helical stacks get closer, the electron $\pi$ orbitals begin to overlap, such that at some point the electron transport pathways, which were initially along the helical backbone of the molecule, now run between stacks. This "through space" transport rather than "through bond" transport means that although the structure is still chiral and even helical, electron transport is not. 

Structural chirality can be easily defined from symmetry \cite{theiler2025non}, but electron chirality seems a more elusive concept. We suggest that a useful working definition for a chiral electronic system is such that an electronic transport system is defined as chiral if  electron transport generates a magnetic field parallel to the integrated or average current. This can also be used to define a "chiral axis" to the electronic system, as discussed in Ref.~\onlinecite{monti2024surface}, where we also discuss ways to measure it. 

\subsubsection{Temperature (T-) dependence of the CISS effect}
While evidence is abundant that the CISS effect decreases with increasing temperature \cite{yang2023,qian2022chiral,rahman2021molecular, rahman2022chirality, hossain2023transverse} (see also references within Ref.~\onlinecite{alwan2023temperature}), there are also indications that in some systems the CISS effect may increase with increasing temperature (many also mentioned in Ref.~\onlinecite{alwan2023temperature}). 

Although there have been some claims that other theoretical approaches can quantitatively describe the T-dependence of the CISS effect, to date the spinterface theory is the only demonstrated approach to do so \cite{yang2023,alwan2023temperature}. For instance, the chiral-phonon model was suggested to account for the T-dependence of the polarization \cite{das2022temperature}, yet a detailed examination of the paper shows that while the T-dependence of the polarization indeed shows agreement between theory and experiment, the raw data show remarkable quantitative difference, so much that the currents themselves decrease with temperature in the theory, while they increase with temperature in the experiments. 

Nonetheless, at least for these systems criticism regarding the T-dependence of the the CISS effect in the spinterface model seems justified, because the spinterface model shows a decrease in the CISS effect with temperature. To address this critique we make two comments: First, we mention that within the transport spinterface model, the CISS effect appears on top of the bare I-V properties of the junction. This means that if the transport properties of the junction are temperature dependent, there is the possibility that the CISS polarization will appear to be temperature-dependent in a way that is not related to the CISS effect itself, but rather to the properties of the system. A clear example is the situation in Ref.~\onlinecite{qian2022chiral}, where a temperature-dependent increase in the total current was erroneously interpreted as an increase in the CISS effect \cite{alwan2023temperature}. 

\begin{figure}
\begin{center}
\includegraphics[width=10.6cm]{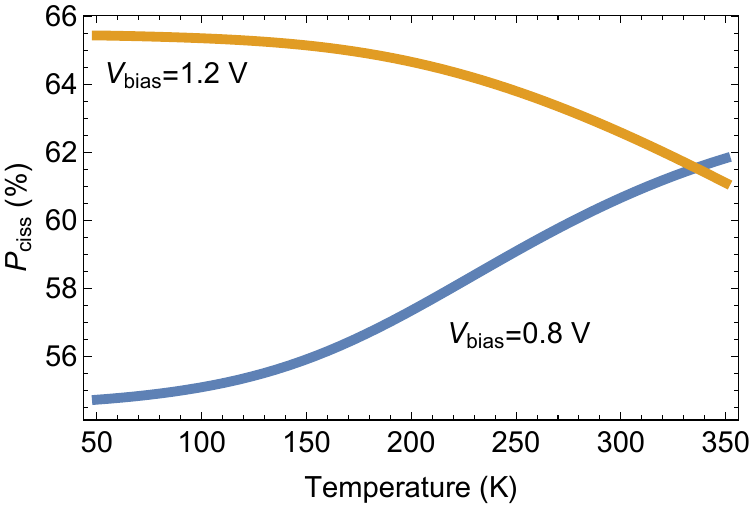}
\caption{Temperature-dependence of the CISS polarization for a single-level dynamical spinterface model (see text for details), at bias voltages $V=0.8V,1.2V$, showing different T-dependence. }
\label{T-dependence_dynamicsspinterface}
\end{center}
\end{figure}

Second, we note that there may be a difference in the T-dependence between the transport spinterface model\cite{alwan2021spinterface} and the \rs{dynamical spinterface model}\cite{monti2024surface}, and within these models the T-dependence may depend on various parameters. To give an example, in Fig.~\ref{T-dependence_dynamicsspinterface} we show the CISS polarization as a function of temperature, evaluated using a single-level transport model within the dynamic spinterface approach. We assume that the bare I-V curve is of the Landauer form of Eq.~\ref{eq:toymodel}, and that the stable surface magnetic moments induce a constant field which shifts the energy levels, $\varepsilon_\si=\varepsilon_0+\si \alpha_A$. The molecular parameters are taken to be $\varepsilon_0=0.6$eV, $\Gamma=0.01$eV, $\alpha_A=0.1$eV. We evaluate the CISS polarization as a function of temperature by calculating the current for the magnetization of the FM electrode pointing parallel and anti-parallel to the current ($s=\pm 1$ in Eq.~\ref{eq:toymodel}), evaluated at bias voltages $V_{\mathrm{bias}}=0.8, 1.2$ V. As can be seen, the CISS polarization exhibits different quantitative T-dependence.

The lesson to be learned from the discussion above and from this example is that the temperature-dependence of the CISS effect may not be universal and in fact may be system-dependent. In deciphering the T-dependence of the CISS effect for a given system it is thus crucial to first understand the T-dependence of the system's transport properties, which may in fact determine the T-dependence of the observed CISS effect in that system.     

\subsubsection{Dependence on electrode SO coupling strength}
While most of the experimental studies of the CISS effect (transport and otherwise) use Au as the metal electrode, there are some exceptions \cite{qian2022chiral,kondou2023spontaneous, gupta2024chirality, bhartiya2024chiral, chae2024promise,vardeny2024manifestation}, which show the CISS effect with other electrodes. As described in Sec.~\ref{The spinterface model}, the original description for the spinterface mechanism \cite{alwan2021spinterface} suggested that the interaction between the localized surface magnetic moments and the electrons incoming to the molecular junction is due to SO interactions in the electrode. Thus, a natural question to ask is what happens when the metallic electrode is changed to a metal or other conducting material that has lower SO coupling? Will this necessarily decrease the magnitude of the CISS effect, e.g., by reducing the CISS polarization? 

Our answer is twofold. First,  as was demonstrated in detail in the appendix of Ref.~\onlinecite{dubi2022spinterface}, an appreciable CISS polarization can be obtained even with very small SO coupling in the electrode. Second, the spinterface model (even the original version) requires two forms of couplings between the surface magnetic moments and electrodes, namely the SO coupling (SOC) and the spin -- exchange interaction (SEI). \rs{By `exchange' we mean two contributions: interfacial s–d exchange between itinerant electrons in the molecule and the localized surface moments of the electrodes, and conventional Heisenberg spin -- spin exchange among the interfacial moments; both participate in the feedback loop -- SOC provides the symmetry breaking, while exchange amplifies and stabilizes the interfacial tilt that sets the CISS barrier.} Both these parameters -- SOC and SEI are crucial in determining the magnitude of the CISS effect. Unfortunately, it is very hard to experimentally differentiate between them, since they are both properties of the specific metal--molecule interface. Thus, even if the SO coupling is reduced, it may be that the spinterface exchange coupling increases, leading to an overall increase in the CISS effect magnitude. 

\subsubsection{CISS without electrodes} Recently, Eckvahl et al. \cite{eckvahl2023direct,eckvahl2024detecting} claimed to observe the CISS effect by probing charge transfer through a donor-chiral bridge-acceptor structure using electron paramagnetic resonance spectroscopy. Since these systems have no metal electrode attached to them, it seems that there is no need for a metal electrode to observe the CISS effect, which raises questions regarding the relevance of the spinterface model \cite{subotnik2023chiral}, since presumably the spinterface model requires the presence of a metal electrode for the CISS effect to develop.

We wish to make two comments regarding this point. The first is that, in fact, the papers by Eckvahl et al. do not unequivocally show the CISS effect. The reason is that in their experiments, the left- and right-handed molecules give essentially {\sl exactly} the same signal, and the control experiment is a non-chiral molecule. Put differently, the symmetry which differentiates between the CISS effect and, say, the magneto-chiral anisotropy effect (as was extensively discussed in Ref.~\onlinecite{rikken2023comparing}), is not resolved in these experiments. Therefore, one should be careful in attributing the results of these experiments to the CISS effect and not to some other chirality-related effect.

\begin{figure}
\begin{center}
\includegraphics[width=10.6cm]{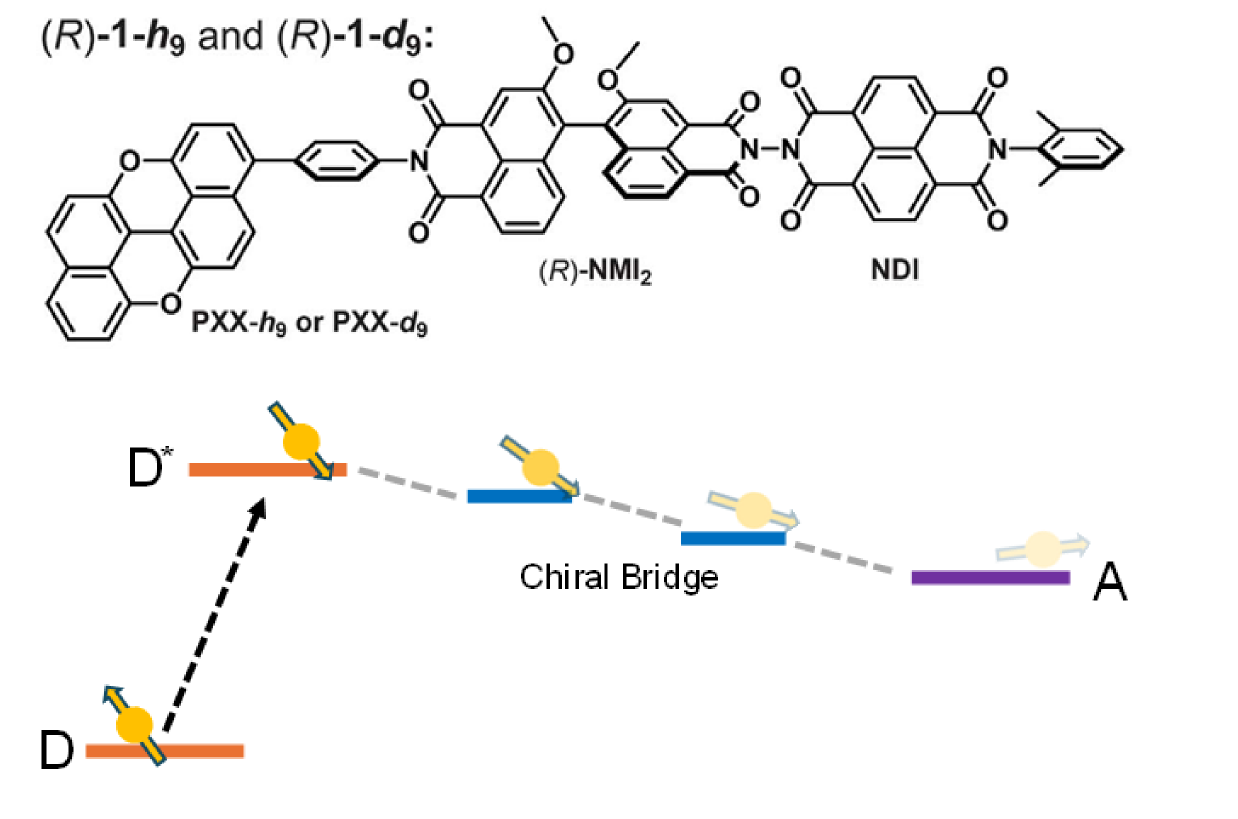}
\caption{Schematic representation of the system under study by Eckvahl et al. \cite{eckvahl2023direct} (top, {adapted with permission from Eckvahl, H. J.; Tcyrulnikov, N. A.; Chiesa, A.; Bradley, J. M.; Young, R. M.; Carretta, S.; Krzyaniak, M. D.; Wasielewski, M. R. Science 2023, \textbf{382}, 197–201. Copyright 2023 The Authors, licensed exclusively to the American Association for the Advancement of Science.}), and a cartoon illustration of the electron transfer mechanism. The immobile electron, which is localized at the donor ground state, can act as a "localized surface moment" to generate the CISS effect. }
\label{DchBA}
\end{center}
\end{figure}

The second comment is that when analyzing the system under study in these experiments, it seems that the spinterface mechanism can actually explain the observations, at least qualitatively at this stage. To explain this, we plot in Fig.~\ref{DchBA} a schematic of the molecular system under study in Ref.~\onlinecite{eckvahl2024detecting}. The molecule is plotted at the top, and the interpretation of the physical process is below. The standard picture is that the donor state ($D$) is doubly occupied at the ground state, and under illumination, one of the electrons is excited (in a $D^*$ state). The excited electron then flows through the chiral bridge to the acceptor state ($A$) where measurement is performed. The point is that during the process of (excited) electron transfer from the excited $D^*$ state to the $A$ state, an immobile electron is always present at the donor ground state. This electron interacts with the excited electron via exchange interaction, therefore acting as an effective "surface magnetic moment", except it is not localized at the surface of an electrode, but rather localized at the donor. We therefore expect that a similar "spinterface" mechanism can operate here to break the spin-symmetry and generate the CISS effect.

\rs{\subsubsection{Where is the time-reversal symmetry breaking coming from?}

It is widely accepted that the Onsager reciprocity relations prohibit spin polarization arising solely from spin-orbit (SO) interactions. Consequently, the CISS effect is thought to require conditions beyond Onsager symmetry, namely the presence of dissipation that breaks time-reversal symmetry (TRS) or operation outside the linear-response regime \cite{evers2022theory, Yang19, Yang2021Linear, tirion2024mechanism, rikken2023comparing}. Experimental data, however, appear to indicate that CISS persists even within the linear-response regime (the readers can judge the data by themselves in, e.g., Fig.~\ref{Buergler_fit}), challenging this theoretical expectation.  

Where, then, does the required TRS breaking originate? The spinterface model does not provide a direct answer. Instead, it assumes dissipation to be present, implemented phenomenologically through the Landau–Lifshitz–Gilbert equation \cite{alwan2021spinterface,monti2024surface}, where dissipation is argued to arise intrinsically from SOC \cite{hickey2009origin}. In recent work \cite{monti2024surface}, it was proposed that TRS is in fact broken as soon as a molecule is brought into contact with a surface. A simple gedanken demonstration illustrates this point: when a molecule attaches to a surface, electrons may flow from the molecule into the electrode’s Fermi sea, producing transient displacement currents until a new equilibrium is reached, in which the molecule becomes charged and a surface dipole forms. Detaching the molecule may not reverse this process - the molecule may remain charged - revealing an inherent irreversibility due to the transient dynamics of charge-transfer at the molecule-surface interface. In this sense, dissipation stems from relaxation processes within the metal. 

Can the microscopic origin of TRS breaking, and thus the apparent experimental deviation from Onsager reciprocity, be established within a rigorous theoretical framework and a microscopic model? We believe the answer is affirmative. Developing such a microscopic theory, however, lies beyond the spinterface model discussed here and is deferred to future work and discussion.  

}

\subsubsection{Brief conclusion from the critical view}
The statistician George Box famously wrote, “All models are wrong, some are useful."\cite{box1976science} A corollary to this statement is that no theory is perfect and problems can be found in any theory. The spinterface approach is no exception to this rule. However, the spinterface model has proven to be quite useful in providing both physical intuition into the CISS effect, as well as a quantitative description of the effect and qualitative explanation to various experimental features (see Sec.~\ref{The spinterface model}). Criticism expressed towards the spinterface model is in many cases justified, and while answers to some points were laid out here, some points are left for the future, as they require additional studies which are beyond the scope of this review. Specifically, a microscopic model for calculating and predicting the coupling parameters of the spinterface model should be developed. 

\subsection{Predictions of the spinterface approach — old and new}\label{predictions}
A key element of any useful theory is its ability to make predictions, and hopefully predictions which can (i) be tested in state-of-the-art experiments, and (ii) can be used to differentiate among competing theories. In previous publications, we have supplied a list of predictions which can readily be tested. Some examples include:   

1.~CISS in a chiral molecule connected to the metal electrode through a non-chiral moiety: We expect that the CISS effect will be reduced as a function of the non-chiral moiety length, because the spinterface exchange coupling is expected to decrease with distance. \cite{alwan2021spinterface}

2.~CISS effect with Ta electrode instead of Au: In Tantalum, the sign of the SO coupling is reversed compared to that of gold. As a result, if indeed the electrode SO coupling is responsible for the CISS effect, we expect that the role of majority and minority spins will be reversed in Ta junctions compared to Au junctions. \cite{alwan2021spinterface} 

3.~Tilting the detector in photoemission CISS experiments: The spinterface theory predicts that the direction of majority and minority spins is determined by the direction of the chiral axis of the molecular layer. This implies that tilting the spin-axis of the detector in photoemission experiments will have a clear signature on the measured CISS polarization. \cite{monti2024surface}



    

In what follows, we provide two new predictions for the CISS effect, which are well within current experimental capabilities, and can provide new insight on the origin of the effect, and importantly, distinguish between the spinterface model and other mechanisms. 

\subsubsection{Transport CISS effect in molecular junctions under illumination}
The key to the experiments we suggest here is that, according to the spinterface model, the CISS signal, which we define as the difference between the I-V curves for two different orientations (parallel and anti-parallel) of the magnetic electrode, is proportional to the bare differential conductance \cite{dubi2022spinterface}. We recall that within the spinterface model, the $J_s(V)$ curves for two different orientations (parallel and anti-parallel, $s=\pm1$) of the magnetic electrode are given by $J_s(V)=J_0(V+s\Lambda(V))$, where $\Lambda(V)$ is the shift in the local chemical potential due to the presence of surface magnetic moments, which can depend on voltage (in the transport spinterface model) or be voltage-independent (for the dynamical spinterface model). In any case, the spin signal is then $\Delta J_{\uparrow \downarrow}=J_\uparrow(V)-J_\downarrow(V)=J_0(V+\Lambda)-J_0(V-\Lambda)\approx 2 G(V) \Lambda$, where $G(V)=J^{'}_0(V)$ is the so-called differential conductance. 

A direct consequence of this simple derivation is that manipulation of a junction's differential conductance $G(V)$ will show up in the CISS signal. In Ref.~\onlinecite{dubi2022spinterface} the CISS signal was successfully compared to the differential conductance (using data from \cite{Xie11}). However, this demonstration was limited by the fact that the data were taken from measurements of different samples in different experiments. A more crucial experiment to demonstrate the connection between the CISS signal and the differential conductance and thus challenge the spinterface mechanism, would be to measure the CISS signal while changing the differential conductance {\sl in situ}. A potential way to achieve that is to illuminate the molecular junction.\cite{tang2022light, sun2024development} 

When a molecular junction is illuminated by off-resonant excitation, i.e. well below the HOMO-LUMO gap of the molecule, the result is typically considered within the framework of the so-called Tien-Gordon (TG) model \cite{tien1963multiphoton,vadai2013plasmon,sun1997influence,chi2012microwave}. For transport through a non-interacting molecular junction, characterized by a transmission function $T(\ve)$ in the dark, the TG model suggests that under illumination, the transmission function becomes
\begin{equation}
T_{TG}(\ve)=\sum_n \bessel_n^2\left(\frac{\Delta}{2}\right) T(\ve+n \hbar \omega)~
\label{Eq:TienGordon}
\end{equation} 
where $\bessel_n(x)$ are Bessel functions of the first kind, $\omega$ is the photon energy from the laser light, and $n$ runs over all integers. $\Delta=E_\omega/\hbar \omega$ is a unitless quantity, where $E_\omega$ is the shift in the electronic energy brought about by illuminating with frequency $\omega$. Simply put, $\Delta$ is a measure for the strength of the laser light illumination. 

As seen from Eq.~\ref{Eq:TienGordon}, the transmission function under illumination develops "side-peaks" at equi-distances $\hbar \omega$ from the resonances of the "dark" transmission function. Since the differential conductance is proportional (approximately, at low temperatures) to the transmission function, $G(V/2)\sim T(V)$ \cite{Book:Cuevas_Scheer10}, it follows that illumination at different wavelengths and different illumination power is a direct way to change the differential conductance of a molecular junction {\sl in situ}. Thus, a direct prediction of the spinterface theory is that under illumination, the CISS signal should change. To make this prediction quantitative, we consider a single resonant level molecular junction characterized by a Lorentzian transmission function \cite{Book:Cuevas_Scheer10},  
\beq
T_{\si s}=\frac{\Gamma^2_{\si s}}{\Gamma^2_{\si s}+(\ve-\ve_0)^2}~~,  \label{eq:Lorntz}
\eeq
where $\ve_0$ is the molecular resonance, and $\Gamma_{\si s}=\frac{1}{2}(\Gamma_L+\Gamma_{R,\si s})$. The total current (per direction $s$ of electrode magnetization) is  
\begin{equation}
    J_{s}(V) = \frac{2e}{h}\sum_{\sigma}\int_{-\infty}^{\infty} d\varepsilon ~T_{\sigma s}(\varepsilon)
    (f_{L}^{\sigma}((\varepsilon), T, V) - f_{R}^{\sigma}(\varepsilon, T, V))~
\label{eq:landauer}\end{equation}
where $f_{L/R}^{\sigma,s} (\varepsilon,T, V)= \left[ 1 +  e^{((\varepsilon-\mu_{\sigma,s}) \pm eV/2)
/(k_{B}T})\right]^{-1} $ are the Fermi-Dirac
distribution of the left (L) and right (R) electrodes, $k_{B}$ is the Boltzmann constant and $T$ the temperature of the electrodes, and $\mu_{\si s}$ are the spin-shifted chemical potentials evaluated from the spinterface theory described in Sec.~\ref{The spinterface model}.

We consider a molecular junction with a resonance at $\ve_0=1.6$ eV (much smaller than a typical HOMO-LUMO gap), a common value for a molecular junction, illuminated by laser light with photon energy $\hbar\omega=1.8$eV, corresponding to standard red laser light (with wavelength 688nm). The TG formula (Eq.~\ref{Eq:TienGordon}) thus predicts that under illumination, the transmission function will develop a side-peak at $\pm (\ve_0-\hbar) \omega=\pm 0.2$eV, which will be observed as a kink in the current (or a peak in the differential conductance) at bias voltage $V=\pm 0.4$ eV.  

\begin{figure}
\begin{center}
\includegraphics[width=10.6cm]{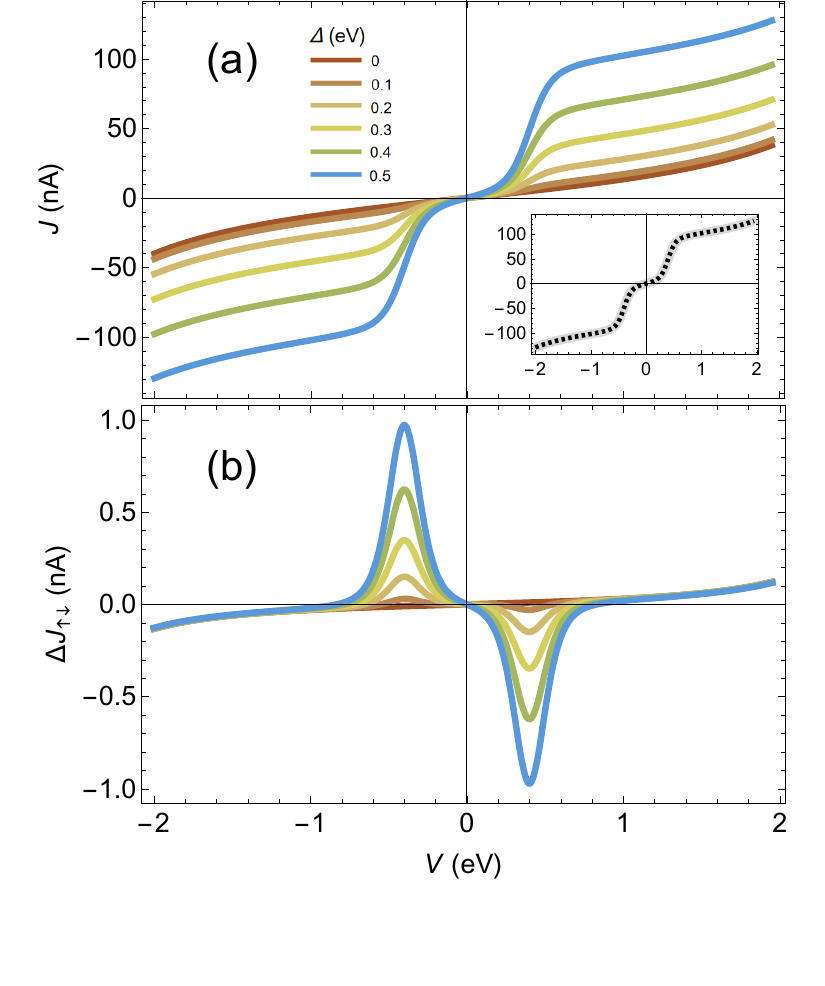}
\caption{(a) Current vs voltage for an illuminated molecular function for different values of $\Delta$ (see text for numerical parameters). Inset:  $J_\uparrow(V)$ and $ J_\downarrow(V)$ (dashed black and solid gray lines, respectively) for $\Delta=0.5$. (b) CISS signal $\Delta J_{\uparrow \downarrow}$ vs voltage for different values of $\Delta$. Under illumination, the CISS signal develops at the resonant illumination energy. }
\label{fig_CISS_Ilum}
\end{center}
\end{figure}

In Fig.~\ref{fig_CISS_Ilum}(a) we plot the bare current vs voltage for the junction under illumination with increasing values of $\Delta$ (Eq.~\ref{Eq:TienGordon}), showing the response of the current to illumination. In the inset of Fig.~\ref{fig_CISS_Ilum}(a) we plot the currents $J_\uparrow(V), J_\downarrow(V)$ (dashed black and solid gray lines, respectively) for $\Delta=0.5$, the differences between them are not visible to the naked eye. 

In Fig.~\ref{fig_CISS_Ilum}(b) we plot the CISS signal, $\Delta J_{\uparrow \downarrow}$, for different values of $\Delta$. As can be seen, the CISS signal shows an enhancement at the resonant energy. While the CISS polarization is only $\sim 1 \%$, this signal should be large enough to be detected within current experimental capabilities.  

\subsubsection{Thermoelectric CISS effect}
Another set of experiments that can distinguish between the spinterface mechanism and other mechanisms for the CISS effect is the so-called thermoelectric CISS effect. The spinterface model assumes that, due to the presence of surface magnetic moments, the energy levels of the molecules, or mathematically equivalent, the chemical potentials, shift in opposite directions for the two electronic spin species. There is an experimental scheme that can test this assumption. 

When a molecular junction is placed between two electrodes kept at different temperatures, a thermoelectric current is generated, due to imbalance between the Fermi distributions on both sides of the junction. Importantly, while energy always flows from the hot electrode to the cold one (thus abiding the second law), the direction of electric current depends on the energy of the frontier orbital. If the orbital energy is above the Fermi energy of the electrodes, then electron current will flow from the hot electrode to the cold one. However, if the orbital energy is below the Fermi level (so-called hole-like transport) then electron current will flow from the cold electrode to the hot one. 

\begin{figure}
\begin{center}
\includegraphics[width=12.6cm]{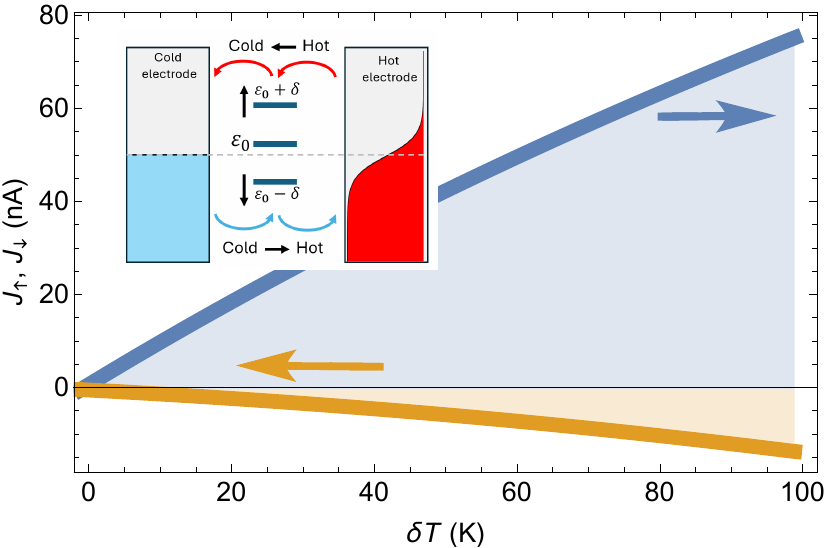}
\caption{Current as a function of temperature-difference for different electrode magnetizations (parallel and antiparallel to the current flow, blue and orange lines respectively). For the relevant junction parameters (see text for details), the current can change direction when the electrode magnetization is reversed (or equivalently the molecular chirality), a signature of the spinterface mechanism. Inset: schematic of the thermo-electric CISS effect. 
}
\label{fig_CISS_thermo}
\end{center}
\end{figure}

Now consider the transport CISS effect within the spinterface model. Due to the presence of stable surface moments, there is a shift in orbital energy in opposite directions for up- and down electrons. Thus, if the frontier orbital is close enough to the Fermi level, and the interaction energy between the surface moments and the electrons is large enough, then a situation may arise that the orbital energy for up-electrons and for down-electrons end up (say) above and below the Fermi energy, respectively. While this would not affect charge current under bias voltage, the associated thermoelectric currents will have {\sl opposite directions} when the magnetic electrode is magnetized parallel or anti-parallel to the direction of current flow. This quantitative description is depicted in the inset of Fig.~\ref{fig_CISS_thermo}.

In the main panel of Fig.~\ref{fig_CISS_thermo}, we plot the (closed circuit thermo-) current for the two magnetization directions (blue and orange lines) as a function of the temperature difference between the electrodes, setting the cold electrode to be at room temperature. We use the single-level Landauer formalism described above (Eq.~\ref{eq:landauer}), with a Lorentzian transmission function. The bias voltage is zero, the level broadening is $\Gamma=10$meV, the magnetic electrode spin-polarization is $30\%$, the orbital energy is at $\varepsilon_0=80$meV, and the spin-dependent shift due to interaction with surface moments is taken to be $\alpha_A=0.1$eV. As can be clearly seen, the currents for the two magnetizations are in opposite directions, a qualitative feature which can readily be tested experimentally. 

Moreover, the same argument suggests that if one could tune the orbital energy of the molecule via gating, then there would be a gate voltage at which the thermoelectric current for one magnetization direction would essentially vanish, because the orbital energy under the (spinterface) CISS effect will be pushed right to the Fermi level. At this value, the CISS polarization of the junction should be $\sim100\%$ for any temperature difference; this can be used to evaluate the interaction energy between the surface moments and the electrons.  

\begin{figure}
\begin{center}
\includegraphics[width=12.6cm]{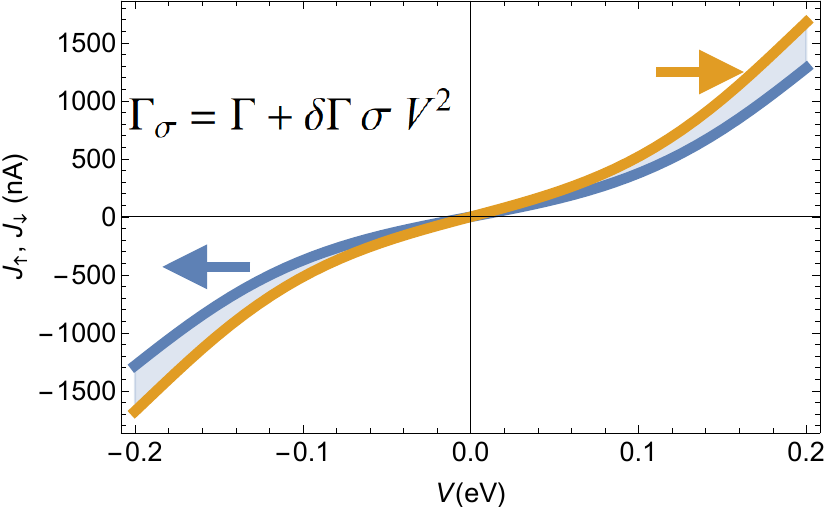}
\caption{ I-V curves for different electrode magnetizations for a model where the molecule-metal electrode coupling is spin-dependent, showing a CISS effect. Numerical parameters are $\Gamma_0=0.01$ eV, $\varepsilon=0.1$ eV, $\delta \Gamma$=5, the Ni electrode polarization is taken to be $\sim 30\%$, and the temperature is room temperature. }
\label{fig_CISS-coupling-dependent}
\end{center}
\end{figure}

Finally, we point out that a thermoelectric CISS experiment can also be used to shed additional light on the origin of the CISS effect, as follows. As we discuss in Sec. \ref{General considerations}, the CISS effect requires breaking of the spin-symmetry (or spin-degeneracy) in the system. In essentially all current models for the CISS effect, the spin-symmetry breaking leads, in some form and through different mechanisms, to lifting the spin-degeneracy in the molecular energy levels. However, in molecular junctions, the coupling strength between the molecule and the electrode plays an equally important role in determining the transport properties. One can easily imagine that interactions such as Coulomb or exchange, through a mechanism similar to the one described in  Ref.~\onlinecite{dubi2013dynamical} will lead to a situation where the coupling between the molecular orbitals and the electrodes becomes spin-dependent. In a single-level Landauer formalism (Eq.~\ref{eq:toymodel}) this would mean having a spin-independent energy level, and having the coupling to the metallic electrode be spin-dependent. In Fig.~\ref{fig_CISS-coupling-dependent} we plot the I-V curves for two electrode magnetizations in a phenomenological model where the coupling is proportional to the voltage squared, i.e. $\Gamma_{L,\sigma}=\Gamma_0 (1+\sigma \delta \Gamma V^2)$, such that the CISS symmetry is preserved. The CISS effect is clearly observed and is essentially indistinguishable from the ``standard" picture of a spin-dependent energy level. 

\begin{figure}
\begin{center}
\includegraphics[width=12.6cm]{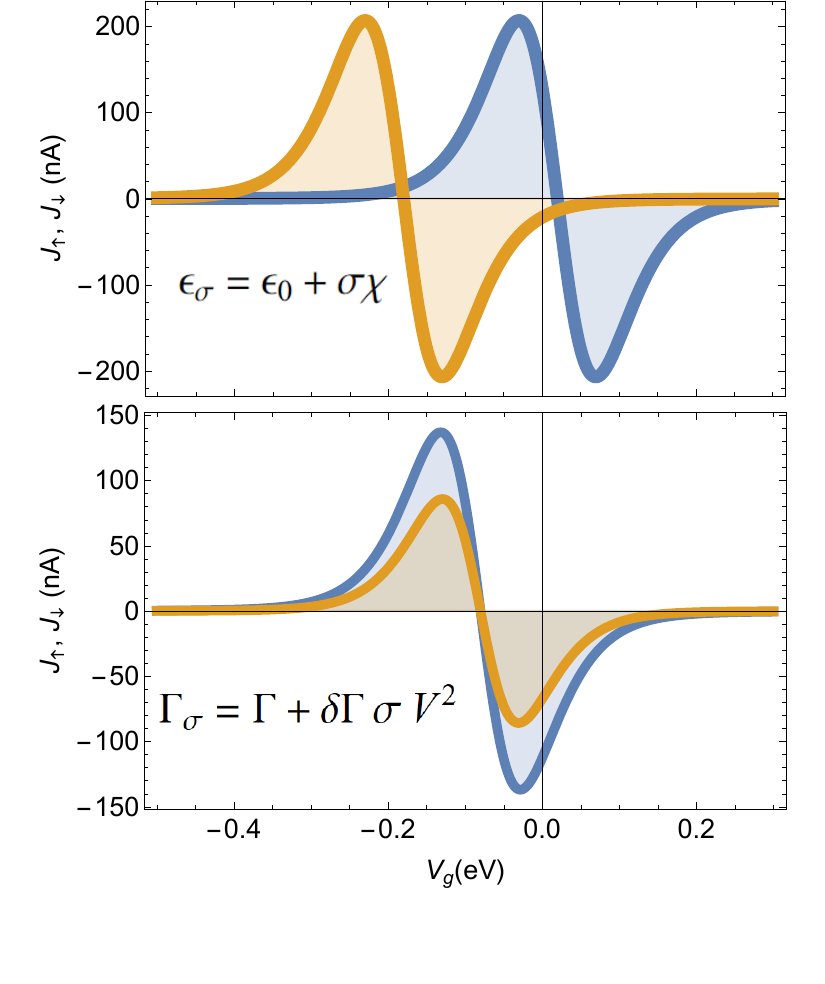}
\caption{{\textbf{Thermoelectric CISS as a mechanism discriminator.} Thermo -- current $J_{\sigma}$ versus gate voltage $V_g$ at fixed $\Delta T = 100\,\mathrm{K}$ for two magnetization directions, with $\sigma = \uparrow$ and $\downarrow$. \textbf{Top:} spin-dependent level-shift model -- $J(V_g)$ curves shift horizontally between magnetizations, enabling extraction of the CISS-related energy shift $\chi$.
\textbf{Bottom:} spin-dependent coupling model -- magnetization reverses the magnitude only; the $V_g$ dependence (including the zero-current crossing) is unchanged.}}
\label{fig_thermoCISS_compare}
\end{center}
\end{figure}

Clearly, although spin-dependent molecular levels or spin-dependent couplings are two distinct mechanisms,  they can both show the CISS effect. A thermoelectric CISS effect is a way to distinguish between them, in a system where the molecular junction can be subjected to a gate voltage~\cite{gehring2019single,chen2024quantum,li2023molecular,kim2022charge,garrigues2016electrostatic}. In Fig.~\ref{fig_thermoCISS_compare} we plot the currents for two electrode polarizations, evaluated vs. a gate voltage, which shifts the molecular orbital, at a constant temperature difference of 100K. The top panel shows the current in a model where the molecular energy level is spin-dependent, and the bottom panel shows a model in which the coupling is spin-dependent (parameters are the same as in Fig.~\ref{fig_CISS-coupling-dependent}). One can clearly see the qualitative difference between the two models. If the CISS effect is due to a shift in the molecular orbital energy, then the gate-dependent current displays a shift in the gate voltage between the two electrode magnetizations, which can also be used to evaluate the CISS-related energy shift $\chi$. On the other hand, if the CISS effect is due to a spin-dependent coupling, then changing the electrode magnetization changes only the magnitude of the current, but no shift in the gate-voltage dependence. Specifically, the point of zero current is the same for both electrode polarizations, a clear indication that there the CISS is not due to some spin-dependent change in the energy levels.  

\section{Summary and concluding thoughts } \label{summary}

In this perspective, we revisited the fundamental experimental observations underlying the chiral-induced spin selectivity (CISS) effect and critically examined the theoretical models proposed to explain it. Experimentally, CISS manifests across a broad spectrum of techniques—from photoemission and charge transport through chiral layers or single molecules to magnetotransport measurements at metal–molecule interfaces. These diverse observations suggest that CISS is not tied to a specific experimental setup but rather arises from a common physical origin or a small set of closely related mechanisms. Most demonstrations of the effect share key universal features: a pronounced dependence on molecular handedness, measurable spin polarization at or above room temperature, and -- so far -- an elusive connection to the specific molecular and chemical details.

These observations raise pressing questions about the mechanisms capable of producing robust spin polarization, given that spin–orbit coupling (SOC) in organic molecules is generally considered too weak -- by orders of magnitude -- to account for the experimentally observed CISS magnitudes. In this work, we evaluated various theoretical frameworks, assessing their strengths and limitations. Among them, we judge the spinterface model to most comprehensively account for the known features of CISS, while also offering testable predictions that have yet to be explored.

The next central challenge is to extend the spinterface framework beyond steady-state transport and phenomenological photoemission models. A successful theory must, for example, account for CISS-polarization in electrode-free electron transfer processes and be capable of predicting time-resolved electron paramagnetic resonance (trEPR) spectra and radical-pair spin chemistry. Given the weakness of atomic SOC in the organic systems where CISS is observed, the essential question becomes: how can interfacial orbital asymmetry between the donor and a chiral bridge, together with environment-induced dephasing, give rise to spin-degeneracy breaking? The interplay between these factors—constructive in some regimes, destructive in others—ultimately determines the achievable degree of spin selectivity.

Advancing the theory will require identifying a small set of dimensionless figures of merit that capture the competition between spin exchange within the donor, electronic coupling across the bridge, long-range exchange interactions, and decoherence times. Moreover, the spinterface framework must be expanded beyond mean-field approximations to leverage chemical and materials design fully. This would enable predictive modeling of systems before synthesis and fabrication. A mature, first-principles spinterface theory could establish general design rules for low-loss spin-polarized sources, chirality-controlled reaction pathways, and chemically addressable quantum bits.

{\bf Acknowledgments:} YD acknowledges support from a BSF grant No.~2023787. OLAM acknowledges support from the U.S. National Science Foundation under grant \#CHE-2513261 and from the Air Force Office of Scientific Research under grant \#FA9550-21-1-0219. AS acknowledges support from the Israel Science Foundation under grant No.~1499\slash 23.

\bibliography{refs}

\end{document}